\documentstyle[12pt]{article}
\def\in{{\rm in}}
\def\out{{\rm out}}
\def\Ci{\mathop{\rm Ci}}
\def\Si{\mathop{\rm Si}}
\def\begineq{\begin{equation}}
\def\endeq{\end{equation}}
\makeatletter
\def\ps@myfrontpage{\let\@mkboth\@gobbletwo
 \def\@oddhead{\large\sl\vbox to0pt{\hbox{Department of Physics}
               \hbox{Kyoto University}\vss}\hfil
              \rm\vbox to0pt{\pubnumber\hbox{\today}\vss}}%
 \def\@oddfoot{}\def\@evenhead{}%
 \def\@evenfoot{}\def\sectionmark##1{}\def\subsectionmark##1{}}
\def\@maketitle{\newpage
\null
\vskip\temp
 \vskip 2em           
 \begin{center}
  {\LARGE \@title \par}     
  \vskip 3em 
  {\large                        
   \lineskip .5em           
   \begin{tabular}[t]{c}\@author
   \end{tabular}\par}
\end{center}
 \par
 \vskip 3em 
}
\makeatother
\def\in{{\rm in}}
\def\out{{\rm out}}
\def\Ci{\mathop{\rm Ci}}
\def\Si{\mathop{\rm Si}}
\def\begineq{\begin{equation}}
\def\endeq{\end{equation}}
\def\pubnumber{\hbox{KUNS 1262}}
\newdimen\temp \temp=0pt
  \setbox1=\vbox{\pubnumber}\advance\temp by 1.5\ht1
\begin{document}

\title{Post-Newtonian expansion of gravitational waves
from a particle in circular orbit around
a Schwarzschild black hole}

\author{Hideyuki Tagoshi and Misao Sasaki\\
{\em Department of Physics, Kyoto University} \\
{\em Kyoto 606-01, Japan}}
\maketitle\thispagestyle{myfrontpage}
\vfill
\begin{abstract}
Based upon the formalism recently developed by one of us (MS),
we analytically perform the post-Newtonian expansion
of gravitational waves from a test particle in circular orbit
of radius $r_0$
around a Schwarzschild black hole of mass $M$.
We calculate gravitational wave forms and luminosity
up to $v^8$ order beyond Newtonian, where $v=(M/r_0)^{1/2}$.
In particular, we give the exact analytical values of the coefficients
of $\ln v$ terms at $v^6$ and $v^8$ orders in the luminosity and
confirm the numerical values obtained previously by the other of us
(HT) and Nakamura. Our result is valid in the small mass limit of one
body and gives an important guideline for the gravitational wave
physics of coalescing compact binaries.
\end{abstract}
\vfill
\newpage

\section{Introduction}
Among the possible sources of gravitational waves, coalescing
compact binaries are the most promising candidates which can
be detected by the near-future laser interferometric
gravitational wave detectors such as LIGO \cite{Ref:ligo}
and VIRGO \cite{Ref:virgo}.
One reason is that we expect such events to occur 3/yr
within 200Mpc \cite{Ref:phinney}.
The other reason is that we expect
enough amplitude of gravitational waves
to be detected by LIGO and VIRGO if such events occur.

Gravitational radiation from coalescing compact binaries contains
rich information about physics of neutron stars,
cosmological parameters, a test of general relativity and so on.
Such information can be extracted out from the gravitational wave form
by the matched filtering technique, that is, by cross-correlating
the incoming noisy signal with theoretical templates.
If the signal and the templates get out of phase with each other by
one cycle as the waves sweep through the LIGO/VIRGO band, their
cross correlation will be significantly reduced.
This means that it is important to construct theoretical
templates
which are accurate to better than one cycle during entire sweep
through the LIGO/VIRGO band \cite{Ref:three}. Thus, much effort
has been recently made to construct accurate theoretical
templates \cite{Ref:Lin,Ref:Wise,Ref:Kidd,Ref:KWW,Ref:wisem}.

To construct theoretical templates,
the post-Newtonian approximations are usually employed to solve
the Einstein equations.
However, based on numerical calculations of the gravitational
radiation from a particle in circular orbit around a
non-rotating black hole, Cutler et al. \cite{Ref:Cutler}
showed that evaluation of the
gravitational wave luminosity to a post-Newtonian order
much higher than presently achieved level
will be required to construct the templates.
Then in order to find out the necessary post-Newtonian order, the
same problem was investigated by Tagoshi and Nakamura \cite{Ref:TN}
with much higher accuracy. They calculated the coefficients of
the post-Newtonian expansion of the gravitational wave luminosity
to $({\rm post})^4$-Newtonian order
(i.e.,$O(v^8)$ beyond Newtonian) and concluded that
the accuracy to at least $({\rm post})^3$-Newtonian order
is required for the construction of theoretical templates.
In addition, they found logarithmic terms in the
luminosity at $({\rm post})^3$ and
$({\rm post})^4$-Newtonian orders.

It is then highly desirable to reproduce these results in
a purely analytical way, that is, to derive the exact analytical
expressions for the coefficients of the post-Newtonian expansion.
Poisson \cite{Ref:poisson} first developed such a method
and calculated the luminosity to $O(v^4)$.
Then extending Poisson's method, a more systematic method was
developed by Sasaki \cite{Ref:sasaki} (hereafter Paper I) and
analytical expressions for the ingoing-wave Regge-Wheeler
functions $X^{\in}_{\ell\omega}$ were derived with the accuracy
required to calculate gravitational wave forms and
luminosity up to $O(v^8)$ beyond Newtonian.
There it was also shown that as long as we are concerned with
the radiation going out to infinity, the effect of the presence of
the black hole horizon does not appear until we calculate to an
extremely higher order, $O(v^{18})$.
Hence the expansion of $X^{\in}_{\ell\omega}$
can be used in a situation in which there is a non-rotating
compact star instead of a black hole.

In this paper, using the result obtained in Paper I,
we calculate analytical expressions for the gravitational wave
forms and luminosity from a particle in circular orbit around
a Schwarzschild black hole to $O(v^8)$ beyond Newtonian.
This paper is organized as follows. In section 2, we
show the general formulae and conventions used in this paper.
In section 3, we first briefly review the result of Paper I
and discuss the relation between the Regge-Wheeler and
Teukolsky functions with emphasis on the consistency
of the post-Newtonian orders in the conversion formula.
In section 4, we present the analytical expressions for the
gravitational wave forms and luminosity to $O(v^8)$ beyond Newtonian
and compare the result with that of Tagoshi and Nakamura \cite{Ref:TN}.
We find an extremely good agreement between the two.
In section 5, applying our result of the luminosity formula,
we discuss the effect of the higher order
post-Newtonian terms to the accumulated phase of gravitational
waves from coalescing compact binaries.
Section 6 is devoted to conclusion.
Throughout the paper we use geometrized units, $c=G=1$.

\section{General formulation}

We consider the case when a test particle of mass $\mu$
travels a circular orbit around a Schwarzschild black hole
of mass $M>>\mu$. We mostly follow notation used by
Poisson \cite{Ref:poisson}, but for definiteness, we
recapitulate necessary formulae and definitions of symbols
in this section.

To calculate the gravitational wave forms and luminosity,
we consider the inhomogeneous Teukolsky
equation \cite{Ref:Teuk,Ref:breuer},
\begin{equation}
\left[{\Delta^2{d\over dr}\left({{1\over \Delta}
{d\over dr}}\right)-U(r)}\right]R_{\ell m\omega}(r)
=T_{\ell m\omega}(r),
\label{eq:teuk}
\end{equation}
where
\begin{equation}
U(r)={r^2\over\Delta}[\omega^2 r^2-4i\omega (r-3M)]-(\ell-1)(\ell+2),
\quad \Delta=r(r-2M),
\end{equation}
and $T_{\ell m\omega}$ is the source term whose explicit form will be
given in Eq.(\ref{eq:tlmw}) below.

We solve Eq.(\ref{eq:teuk}) by the Green function method.
For this purpose, we need a homogeneous solution
$R^{\in}_{\ell\omega}(r)$ of Eq.(\ref{eq:teuk})
which satisfies the following boundary condition,
\begin{equation}
R^{\in}_{\ell\omega}(r)=
  \cases{D_{\ell\omega}\Delta^2 e^{-i\omega r^*} &for~
$r^*\rightarrow -\infty$ ,\cr
         r^3 B^{\out}_{\ell\omega}e^{i\omega r^*}
        +r^{-1}B^{\in}_{\ell\omega}e^{-i\omega r^*} &for~
$r^*\rightarrow +\infty$~,\cr}
\label{eq:boundary}
\end{equation}
where $r^*=r+2M\ln(r/2M-1)$. Then the outgoing-wave solution of
Eq.(\ref{eq:teuk}) at infinity with the appropriate boundary condition
at horizon is given by
\begin{eqnarray}
R_{\ell m \omega}(r\rightarrow\infty)
&=&{r^3e^{i\omega r^*}\over 2i\omega B^{\in}_{\ell\omega}}
\int^{\infty}_{2M}dr R^{\in}_{\ell\omega} T_{\ell m \omega}(r)
\Delta^{-2}
\nonumber\\
&\equiv&r^3 e^{i\omega r^*}\tilde{Z}_{\ell m \omega}.
\label{eq:rlmw}
\end{eqnarray}
In the case of a circular orbit,
the specific energy $\tilde E$ and
angular momentum $\tilde L$
of the particle are given by
\begineq
\tilde E =(r_0-2M)/\sqrt{r_0(r_0-3M)},
\label{eq:ene}
\endeq
and
\begineq
\tilde L =\sqrt{Mr_0}/\sqrt{1-3M/r_0},
\endeq
where $r_0$ is the orbital radius.
The angular frequency is given by $\Omega$ $=(M/r_0^3)^{1/2}$.
Then $T_{\ell m\omega}(r)$ is given by
\begin{eqnarray}
(T_{\ell m\omega}/\pi)/\mu
&=&\biggl\{ -2 _0b_{\ell m}(r_0-2M)^2\delta(r-r_0)
\nonumber\\
& &-_{-1}b_{\ell m}2ir_0\left[{(r_0-2M)^2\delta'(r-r_0)
-(r_0-2M)(2-i\omega r_0)\delta(r-r_0)}\right]
\nonumber\\
& & +_{-2}b_{\ell m} \biggl[r_0^2(r_0-2M)^2\delta''(r-r_0)
\nonumber\\
& &+\{2i\omega r^3_0(r_0-2M)-2r_0(3r_0^2-8r_0M+4M^2)\}\delta'(r-r_0)
\nonumber\\
& &+\left\{{4r_0^2-8M^2-\omega^2r_0^4-6i\omega r_0^2(r_0-M)
}\right\}\delta(r-r_0) \biggr]\biggr\}\delta(\omega-m\Omega),
\label{eq:tlmw}
\end{eqnarray}
where $\delta(r)$ is the Dirac delta function and $'=d/dr$.
The coefficients ${}_sb_{\ell m}$ are given by
\begin{eqnarray}
_0b_{\ell m}&=&{1\over 2}[(\ell-1)\ell(\ell+1)(\ell+2)]^{1/2}\,
_0Y_{\ell m}\left({{\pi\over 2},0}\right)\tilde Er_0/(r_0-2M),
\nonumber\\
_{-1}b_{\ell m}&=&[(\ell-1)(\ell+2)]^{1/2}~
_{-1}Y_{\ell m}\left({{\pi\over 2},0}\right)\tilde L/r_0,
\nonumber\\
_{-2}b_{\ell m}&=&\,_{-1}Y_{\ell m}\left({{\pi\over 2},0}\right)
\tilde L\Omega,
\label{eq:defb}
\end{eqnarray}
where $_sY_{\ell m}(\theta,\varphi)$ are the
spin-weighted spherical harmonics.
 From Eqs.(\ref{eq:rlmw}) and (\ref{eq:tlmw}), we see that
$\tilde{Z}_{\ell m \omega}$ takes the form,
\begineq
\tilde{Z}_{\ell m \omega}=Z_{\ell m}\delta(\omega-m\Omega),
\endeq
where
\begin{eqnarray}
Z_{\ell m}
={\pi\over2i\omega r_0^2B^{\in}_{\ell\omega}}\Biggl\{
&&\Biggl[-_0b_{\ell m}
-2i\,_{-1}b_{\ell m}\left(1+{i\over 2}\omega r_0^2/(r_0-2M)\right)
\nonumber\\
&&+i\,_{-2}b_{\ell m}\omega r_0(1-2M/r_0)^{-2}
\left(1-M/r_0+{1\over 2}i\omega r_0\right)\Biggr]
R^{\in}_{\ell\omega}(r_0)
\nonumber\\
&&+\Biggl[i\,_{-1}b_{\ell m}-\,_{-2}b_{\ell m}
\left(1+{i\over 2}\omega r_0^2/(r_0-2M)\right)\Biggr]
r_0 R^{\in}_{\ell\omega}{}'(r_0)
\nonumber\\
&&+{1\over 2}\,_{-2}b_{\ell m}{r_0}^2 R^{\in}_{\ell\omega}{}''(r_0)
\Biggr\}.
\label{eq:zlmw}
\end{eqnarray}
We note the symmetry, $Z_{\ell, -m}$ $=(-1)^\ell \bar{Z}_{\ell m}$,
which may be seen from Eqs.(\ref{eq:defb}), (\ref{eq:zlmw}) and
the symmetry of the spin weighted spherical harmonics,
$\,_sY_{\ell m}(\pi/2,0)$ $=(-1)^{(s+\ell)}\,_{s}Y_{\ell m}(\pi/2,0)$.
In terms of the amplitudes $Z_{\ell m}$, the gravitational wave form
at infinity is given by
\begineq
h_{+}-ih_{\times}=-{2\mu\over r}\sum_{\ell m}{1\over \omega^2}
{Z_{\ell m}}\,_{-2}Y_{\ell m}(\theta,\varphi)e^{-i\omega(t-r^*)},
\label{eq:wave}
\endeq
and the luminosity is given by
\begineq
{dE\over dt}=\sum_{\scriptstyle \ell=2}^{\infty}\sum_{m=1}^\ell
\mid Z_{\ell m}\mid^2/2\pi\omega^2,
\label{eq:lum}
\endeq
where $\omega=m\Omega$.
Thus the only remaining task is to calculate the ingoing-wave
Teukolsky function $R^{\in}_{\ell\omega}$.

In stead of directly calculating $R^{\in}_{\ell\omega}$ from
the homogeneous Teukolsky equation,
it is much easier to calculate the corresponidng Regge-Wheeler
function $X^{\in}_{\ell\omega}$ first and then transform it
to $R^{\in}_{\ell\omega}$.
The homogeneous Regge-Wheeler equation takes the form \cite{Ref:RW},
\begineq
\left[{{d^2\over dr^{*2}}+\omega^2-V(r)}\right]X_{\ell\omega}(r)
=0, \label{eq:rw}
\endeq
where
\begineq
V(r)=\left({1-{2M\over r}}\right)\left({{\ell(\ell+1)\over r^2}
-{6M\over r^3} }\right).
\endeq
It is known that this equation is obtained by
transforming $R_{\ell\omega}$ as \cite{Ref:Chan,Ref:SN}
\begineq
R_{\ell m\omega}=\Delta\left({{d \over dr^*}+i\omega}\right)
{r^2\over \Delta}\left({{d \over dr^*}+i\omega}\right)r
X_{\ell\omega}. \label{eq:conv}
\endeq
Conversely, we can express $X_{\ell\omega}$ in terms of
$R_{\ell\omega}$ by the inverse transformation formula as
\begineq
X_{\ell\omega}={r^5\over c_0\Delta}
          \left({d \over dr^*}-i\omega\right){r^2\over\Delta}
          \left({d \over dr^*}-i\omega\right)
               {R_{\ell\omega}\over r^2},
\label{eq:inconv}
\endeq
where $c_0=(\ell-1)\ell(\ell+1)(\ell+2)-12iM\omega$.
Then we obtain the asymptotic forms of $X^{\in}_{\ell\omega}$ as
\begin{equation}
X^{\in}_{\ell\omega}(r)=\cases{C_{\ell\omega}e^{-i\omega r^*},
& $r^*\rightarrow -\infty$ \cr
A^{\out}_{\ell\omega}e^{i\omega r^*}
+A^{\in}_{\ell\omega}e^{-i\omega r^*},
 & $r^*\rightarrow +\infty$. \cr}
\label{eq:bou}
\end{equation}
where $A^{\in}_{\ell\omega}$, $A^{\out}_{\ell\omega}$
and $C_{\ell\omega}$ are respectively related to
$B^{\in}_{\ell\omega}$, $B^{\out}_{\ell\omega}$ and
$D_{\ell\omega}$ defined in Eq.(\ref{eq:boundary}) as
\begin{eqnarray}
B^{\in}_{\ell\omega}&=&
 -{c_0\over4\omega^2}A^{\in}_{\ell\omega},
\nonumber\\
B^{\out}_{\ell\omega}&=&
 -4\omega^2 A^{\out}_{\ell\omega},
\nonumber\\
D_{\ell\omega}&=&
  {c_0\over16(1-2iM\omega)(1-4iM\omega)}C_{\ell\omega}.
\label{eq:bin}
\end{eqnarray}
%

\section{Post-Newtonian expansion of the Teukolsky function}
\subsection{Method}

In Paper I, the post-Newtonian expansion of the ingoing-wave
Regge-Wheeler functions $X^{\in}_{\ell\omega}$ was formulated and
they were calculated to $O(\epsilon^2)$ where
$\epsilon\equiv2M\omega$.
In this subsection we briefly review the method.

We rewrite the homogeneous Regge-Wheeler equation as
\begin{equation}
\left[{{d^2\over dz^{*2}}
+1-\left({1-{\epsilon\over z}}\right)\left({{\ell(\ell+1)\over z^2}
-{3\epsilon\over z^3} }\right) }\right]X^{\in}_{\ell}
=0, \label{eq:newrw}
\end{equation}
where $z=wr$, $z^*=z+\epsilon\ln(z-\epsilon)$
and we have suppressed the index $\omega$ since it is trivially
absorbed in $\epsilon$ and $z$.
Note that for $\omega=m\Omega$, $\epsilon=2mv^3$ and $z=mv$ at $r=r_0$.
Hence the post-Newtonian expansion corresponds to expanding
 $X^{\in}_\ell$ with respect to $\epsilon$ and evaluating
$X^{\in}_\ell$ at $z\ll1$ as well as $A^{\in}_\ell$ to required
orders in $\epsilon$.
Now setting
\begin{equation}
X_{\ell}^{\in}=e^{-i\epsilon\ln(z-\epsilon)}z\xi_{\ell}(z),
\label{eq:defxsi}
\end{equation}
we find that Eq.(\ref{eq:newrw}) becomes
\begin{equation}
\left[{d\over dz^2}+{2\over z}{d\over dz}
+\left(1-{\ell(\ell+1)\over z^2}\right)\right]\xi_{\ell}
=\epsilon e^{-iz}{d\over dz}\left[{1\over z^3}{d\over dz}
\left(e^{iz}z^2\xi_{\ell}(z)\right)\right].
\label{eq:rw2}
\end{equation}
Thus expanding $\xi_{\ell}$ with respect to $\epsilon$ as
\begin{equation}
\xi_{\ell}(z)=\sum_{n=0}^\infty \epsilon^n \xi^{(n)}_{\ell}(z),
\label{eq:expan}
\end{equation}
we obtain the recursive equations,
\begin{equation}
\left[{d\over dz^2}+{2\over z}{d\over dz}
+\left(1-{\ell(\ell+1)\over z^2}\right)\right]\xi_{\ell}^{(n)}(z)
=\epsilon e^{-iz}{d\over dz}\left[{1\over z^3}{d\over dz}
\left(e^{iz}z^2\xi_{\ell}^{(n-1)}(z)\right)\right].
\label{eq:rec}
\end{equation}

First for $n=0$, we have
\begineq
\xi^{(0)}_\ell=\alpha^{(0)}j_\ell+\beta^{(0)}n_\ell,
\endeq
where $j_\ell$ and $n_\ell$ are the usual spherical Bessel functions.
The boundary condition is that $\xi^{(0)}_\ell$ be regular
at $z=0$ \cite{Ref:poisson,Ref:sasaki}. Hence $\beta^{(0)}=0$
and for convenience we set $\alpha^{(0)}_\ell=1$.
For $n\geq1$, we rewrite Eq.(\ref{eq:rec})
in the indefinite integral form,
\begin{eqnarray}
\xi^{(n)}_{\ell}&=&n_{\ell}\int^z dz z^2 e^{-iz}j_{\ell}
\left[{1\over z^3}\left(e^{iz}z^2\xi^{(n-1)}(z)\right)'\right]'
\nonumber\\
& &-j_{\ell}\int^z dz z^2 e^{-iz}n_{\ell}
\left[{1\over z^3}\left(e^{iz}z^2\xi^{(n-1)}(z)\right)'\right]'.
\label{eq:xi}
\end{eqnarray}
If the above indefinite integrals can be explicitly performed, it is
easy to obtain $\xi^{(n)}_\ell$ with a desired boundary condition.
In Paper I, it is shown that this is indeed the case for $n=1$ and $2$,
and the boundary condition is that $\xi^{(n)}_\ell$ be also regular
at $z=0$ at least for $n\leq3$.
We then find for $n=1$,
\begin{eqnarray}
\xi^{(1)}_\ell &=&{(\ell-1)(\ell+3)\over 2(\ell+1)(2\ell+1)}j_{\ell+1}
-\left({{\ell^2-4}\over 2\ell(2\ell+1)}
+{{2\ell-1}\over \ell(\ell-1)}\right)j_{\ell-1} \nonumber\\
 & &+z^2(n_\ell j_0-j_\ell n_0)j_0
+\sum_{k=1}^{\ell-2}\left({1\over k}+{1\over k+1}\right)
z^2(n_\ell j_k-j_\ell n_k)j_k \nonumber\\
 & &+n_\ell({\Ci} 2z-\gamma-\ln 2z )-j_\ell{\Si}2z +ij_\ell\ln z
+\alpha^{(1)}_{\ell}j_\ell,
\label{eq:xsi}
\end{eqnarray}
where $\Ci(x)=-\int^{\infty}_x dt\cos t/t$ and
$\Si(x)=\int^x_0 dt\sin t/t$ are the cosine and sine integral
functions, respectively, and $\alpha^{(1)}_\ell$ is an arbitrary
integration constant which represents the
arbitrariness of the normalization of $X_\ell^{\in}$. We set
$\alpha^{(1)}_\ell=0$. As for $n=2$, closed analytical expressions of
$\xi^{(2)}_\ell$ for $\ell=2$, $3$ and $4$ can be found in Paper I,
\S4.2.

To obtain $X^{\in}_\ell$ from $\xi_\ell$ is straightforward.
Decomposing the real and imaginary parts of $\xi_{\ell}^{(n)}$ as
\begin{equation}
\xi_{\ell}^{(n)}=f^{(n)}_\ell+ig^{(n)}_\ell,
\label{eq:fg}
\end{equation}
we find the imaginary parts $g^{(n)}_\ell$ are given by
\begin{equation}
g^{(0)}=0,\quad g^{(1)}=j_\ell\ln z,\quad
g^{(2)}_\ell=-{1\over z}j_\ell+f^{(1)}_\ell\ln z.
\end{equation}
These relations confirm that $X^{\in}_\ell$ is real at least up to
$O(\epsilon^2)$. In fact, inserting Eq.(\ref{eq:fg}) into
Eq.(\ref{eq:defxsi}) and expanding the result with respect to
$\epsilon$ by assuming $z\gg \epsilon$, we find
\begin{eqnarray}
X^{\in}_\ell&=&e^{-i\epsilon\ln(z-\epsilon)}z
\left(j_\ell+\epsilon(f^{(1)}_\ell+ig^{(1)}_\ell)
+\epsilon^2(f^{(2)}_\ell+ig^{(2)}_\ell)+\ldots \right)\nonumber\\
&=&z\left(j_\ell+\epsilon f^{(1)}_\ell
+\epsilon^2
 \left(f^{(2)}_\ell+g^{(1)}_\ell\ln z-{1\over 2}j_\ell(\ln z)^2\right)
+\ldots\right)\nonumber\\
&=&z\left(j_\ell+\epsilon f^{(1)}_\ell
+\epsilon^2
 \left(f^{(2)}_\ell+{1\over 2}j_\ell(\ln z)^2\right)+\ldots\right).
\end{eqnarray}

\subsection{Calculation}

Once we have closed analytical expressions of $\xi^{\in}_\ell$,
$A^{\in}_\ell$ can be readily obtained by gathering all the
coefficients of leading terms proportional to $e^{-iz}/z$ from
the asymptotic forms of $\xi_\ell$ at $z=\infty$.
The explicit forms of $A^{\in}_\ell$ from $\ell=2$ to 4 which are
correct up to $\epsilon^2$ order are given by
\begin{eqnarray}
A^{\in}_2&=&-{1\over 2}ie^{-i\epsilon(\ln 2\epsilon+\gamma)}
\exp{\left(i\epsilon{5\over 3}-i\epsilon^2{107\over 420}\pi\right)}
\nonumber\\
 &\times&\left(1-\epsilon{\pi\over 2}
+\epsilon^2\left\{{25\over 18}+{5\over 24}\pi^2
+{107\over 210}(\gamma+\ln 2)\right\}+\ldots \right),
\label{eq:ain2}
\end{eqnarray}
\begin{eqnarray}
A^{\in}_3&=&{1\over 2}ie^{-i\epsilon(\ln 2\epsilon+\gamma)}
\exp{\left(i\epsilon{13\over 6}-i\epsilon^2{13\over 84}\pi\right)}
\nonumber\\
 &\times&\left(1-\epsilon{\pi\over 2}
+\epsilon^2\left\{{169\over 72}+{5\over 24}\pi^2
+{13\over 42}(\gamma+\ln 2)\right\}+\ldots \right),
\end{eqnarray}
\begin{eqnarray}
A^{\in}_4&=&{1\over 2}ie^{-i\epsilon(\ln 2\epsilon+\gamma)}
\exp{\left(i\epsilon{149\over 60}-i\epsilon^2{1571\over 13860}
\pi\right)}\nonumber\\
 &\times&\left(1-\epsilon{\pi\over 2}
+\epsilon^2\left\{{22201\over 7200}+{5\over 24}\pi^2
+{1571\over 6930}(\gamma+\ln 2)\right\}+\ldots \right),
\label{eq:ain4}
\end{eqnarray}
For $\ell\geq5$, explicit forms of $A^{\in}_{\ell}$ which
are valid up to order $O(\epsilon)$ are given by
\begineq
A^{\in}_{\ell}={1\over 2}i^{\ell+1}
e^{-i\epsilon(\ln 2\epsilon+\gamma)}
\left[1-\epsilon{\pi\over 2}+{i\epsilon\over 2}
\left\{ \sum_{k=1}^{\ell-1}{1\over k}
+\sum_{k=1}^{\ell}{1\over k}+{(\ell-1)(\ell+3)\over \ell(\ell+1)}
\right\} \right].
\label{eq:ain}
\endeq
As discussed in Paper I, and will be mentioned below,
the above formulae are what we all need for $A^{\in}_\ell$ in order
to calculate the wave forms and luminosity to $O(v^8)$
beyond Newtonian.

In addition to $A^{\in}_\ell$, we need the series expansion formulae
of $X^{\in}_\ell$ at $z\ll1$. To clarify their required orders
of accuracy, we first consider the qualitative behavior of
$X^{\in}_\ell$ as $z\rightarrow0$. We find
\begin{eqnarray}
X^{(\in)}_2&=&z^3\left[O(1)+\epsilon O(z)
+\epsilon^2 \{O(1)+O(1)\ln z\}+\epsilon^3 O(z^{-1})+\ldots
\right], \nonumber\\
X^{(\in)}_3&=&z^3\left[O(z)+\epsilon O(1)
+\epsilon^2 \{O(z)+O(z)\ln z\}+\ldots \right], \nonumber\\
X^{(\in)}_\ell &=&z^3\left[O(z^{\ell-2})
+\epsilon O(z^{\ell-3})
+\epsilon^2 \{O(z^{\ell-4})+O(z^{\ell-2})\ln z\}+\ldots \right]
{}~(\ell\geq 4). \label{eq:asyxin}\\
\end{eqnarray}
The above result tells us the accuracy of $X^{\in}_\ell$
needed to achieve $O(v^8)$ beyond Newtonian.
For convenience, we set
\begineq
X^{\in}_{\ell}=\sum_{n=0}^{\infty}\epsilon^n X^{(n)}_{\ell}.
\label{eq:exxin}
\endeq
First, note that we must calculate $X^{(0)}_{\ell}$ to $O(z^{11})$,
$X^{(1)}_{\ell}$ to $O(z^8)$, $X^{(2)}_{\ell}$ to $O(z^5)$
and $X^{(3)}_{\ell}$ to $O(z^2)$.
Then we see from Eq.(\ref{eq:exxin}) that
we need the series expansions of $X^{\in}_\ell$ at $z=0$
to the above required orders for $\ell=2$ to $O(\epsilon^3)$,
 for $\ell=3$ 4, 5 and 6 to
$O(\epsilon^2)$, for $\ell=7$ and 8 to $O(\epsilon^1)$, and
for $\ell=9$ and 10 to $O(\epsilon^0)$.

Since we have closed analytical expressions of $X^{(0)}_\ell$
and $X^{(1)}_\ell$ for all $\ell$, and $X^{(2)}_\ell$ for $\ell\leq4$,
whose series expansion formulae can be trivially obtained,
what we additionally need to do is to derive the series expansions
of $X^{(3)}_2$, $X^{(2)}_5$ and $X^{(2)}_6$.
Taking account of the required accuracy mentioned above,
we find that only the leading terms in their series expansions
are necessary.
First consider $X^{(2)}_5$ and $X^{(2)}_6$.
With the bondary condition that these be regular at $z=0$,
and using Eq.(\ref{eq:xi}) with $n=2$ and Eq.(\ref{eq:xsi}),
They can be easily calculated. We find
\begin{eqnarray}
X^{(2)}_5&=&{{{z^4}}\over {7425}} +O(z^6),
\\
X^{(2)}_6&=&{4\over 212355} z^5+O(z^7),
\end{eqnarray}
where possible ambiguities due to the choice of integration
constants appear only at $O(z^6)$ and $O(z^7)$, respectively,
hence do not affect the coefficients of the leading terms,
as it should be so.
In the same way, the leading behavior of $X^{(3)}_2$ at $z=0$
is found to be
\begineq
X^{(3)}_2={319\over6300}z^2+O(z^4),
\endeq
where the aforementioned ambiguity appears at $O(z^4)$, hence again
does not affect the result.
In Appendix A, we show the series expansions of $X^{(n)}_\ell$
up to the required orders.

To calculate gravitational waves from the Teukolsky equation,
we need to know $R^{\in}_\ell(z)$ at $z\ll1$, which we calculate
by using the conversion formula from $X^{\in}_\ell$ to
$R^{\in}_\ell$, Eq.(\ref{eq:conv}).
In terms of $X^{(n)}_\ell$, the result is expressed as
\begin{eqnarray}
\omega R^{\in}_\ell
&=&z(z-\epsilon)\left({{d \over dz^*}+i}\right)
{z^2\over z(z-\epsilon)}\left({{d \over dz^*}+i}\right)z
X_{\ell}^{\in}(z) \nonumber\\
&=&-{z^2}\,(z {X_{\ell}^{(0)}})
+ {z^2}\,(z {X_{\ell}^{(0)}})''
+ 2\,i\,{z^2}\,(z {X_{\ell}^{(0)}})'
\nonumber\\
&+& {\epsilon}\, \Bigl(
- 2\,z\,(z {X_{\ell}^{(0)}})''
+ {z^2}\,(z {X_{\ell}^{(1)}})''
- {z^2}\,(z {X_{\ell}^{(1)}})
\nonumber\\
& &-2\,i\,z\,(z {X_{\ell}^{(0)}})'
- i\,(z {X_{\ell}^{(0)}})
+ 2\,i\,{z^2}\,(z {X_{\ell}^{(1)}})'
 \Bigr)  \nonumber\\
&+& {{{\epsilon}}^2}\,\Bigl(
+ (z {X_{\ell}^{(0)}})''
- 2\,z\,(z {X_{\ell}^{(1)}})''
+ {z^2}\,(z {X_{\ell}^{(2)}})''
- {z^2}\,(z {X_{\ell}^{(2)}})
\nonumber\\
& &
-i\,(z {X_{\ell}^{(1)}})
-2\,i\,z\,(z {X_{\ell}^{(1)}})'
+2\,i\,{z^2}\,(z {X_{\ell}^{(2)}})'
\Bigr)
\nonumber\\
&+&{{{\epsilon}}^3}\,\Bigl(
+ (z {X_{\ell}^{(1)}})''
-  2\,z\,(z {X_{\ell}^{(2)}})''
+ {z^2}\,(z {X_{\ell}^{(3)}})''
- {z^2}\,(z {X_{\ell}^{(3)}})
\nonumber\\
& &
-i\,(z {X_{\ell}^{(2)}})
- 2\,i\,z\,(z {X_{\ell}^{(2)}})'
+2\,i\,{z^2}\,(z {X_{\ell}^{(3)}})'
\Bigr),
\label{eq:wrin}
\end{eqnarray}
where $'=d/dz$.

Here a problem might arize, since what we have calculated are
the series expansions of $X^{\in}_\ell(z_0)$ to $O(v^8)$
relative to the leading order, but what we actually need are
those of $R^{\in}_\ell$.
Hence we must examine if there is consistency of the post-Newtonian
orders between the series expansions of
$X^{\in}_\ell$ and $R^{\in}_\ell$.

A straightforward way to check this consistency is to insert
the truncated series expressions of $X^{(n)}_\ell$ accurate to
$O(v^8)$ beyond Newtonian to Eq.(\ref{eq:wrin}),
insert the resulting expression of $R^{\in}_\ell$ to
the inverse transformation formula (\ref{eq:inconv}), and examine if
the final result is the same as the starting
expression of $X^{\in}_\ell$.
We have verified this is indeed the case.

As for $B^{\in}_{\ell}$, it should be now obvious
from Eqs.(\ref{eq:wave}), (\ref{eq:lum}) and (\ref{eq:bin})
that the calculated $A^{\in}_\ell$ have sufficient accuracy
to obtain $B^{\in}_\ell$ to the required orders.

All these facts support our confidence that
$X^{\in}_\ell$ and $R^{\in}_\ell$ are equivalent also within
the framework of our post-Newtonian expansion.

\section{Wave forms and luminosity to $O(v^8)$}

Let us now calculate the gravitational wave forms and
luminosity up to $O(v^8)$ beyond Newtonian.
The task is straightforward but tedious.
So we only show the key equations.
The calculations have been performed with the help of the
 computer manipulation software {\em Mathematica}.
We follow the notation of Poisson \cite{Ref:poisson}
to describe the post-Newtonian expansion of gravitational wave
forms and luminosity.
Namely, we express them as
\begineq
{dE\over dt}={32\over 5}\left({\mu\over M}\right)^2
\left({M\over r_0}\right)^5
\sum_{\ell=2}^{\infty}\sum_{m=1}^{\infty}\eta_{\ell m},
\endeq
\begineq
h_{\ell m}^{+,\times} + h_{\ell,-m}^{+,\times}
=-\left({\mu\over r}\right)\left({M\over r_0}\right)
\zeta_{\ell m}^{+,\times}.
\label{eq:zetadef}
\endeq

\subsection{Luminosity}

The procedure is straightforward.
First we obtain $R^{\in}_\ell$ from Eq.(\ref{eq:wrin})
and $B^{\in}_\ell$ from Eq.(\ref{eq:bin}).
Then we insert them to Eq.(\ref{eq:zlmw}) and set
$z=mv$ and $\epsilon=2mv^3$ to obtain $Z_{\ell m}$.
Finally, inserting them to Eq.(\ref{eq:lum}) and expanding
the results with respect to $v$, we obtain $\eta_{\ell m}$.

The explicit forms of $\eta_{\ell m}$ are given by
\begin{eqnarray*}
\lefteqn{\eta_{2,2}=
1 - {{107\,{v^2}}\over {21}} + 4\,\pi \,{v^3}
+ {{4784\,{v^4}}\over {1323}} -
  {{428\,\pi \,{v^5}}\over {21}} + {{19136\,\pi \,{v^7}}\over {1323}}
}\\
& &
+ {v^6}\,\left( {{99210071}\over {1091475}}
- {{1712\,\gamma }\over {105}}
+ {{16\,{{\pi }^2}}\over 3} - {{3424\,\ln 2}\over {105}} -
     {{1712\,\ln v}\over {105}} \right)
\\
& &
+  {v^8}\,\biggl( -{{27956920577}\over {81265275}} +
   {{183184\,\gamma }\over {2205}} - {{1712\,{{\pi }^2}}\over {63}}+
  {{366368\,\ln 2}\over {2205}} + {{183184\,\ln v}\over {2205}}
\biggr),
\\
\lefteqn{\eta_{2,1}=
{{{v^2}}\over {36}} - {{17\,{v^4}}\over {504}}
+ {{\pi \,{v^5}}\over {18}} -
  {{2215\,{v^6}}\over {254016}} - {{17\,\pi \,{v^7}}\over {252}}
}\\
& &
+{v^8}\,\left( {{15707221}\over {26195400}}
- {{107\,\gamma }\over {945}} +
     {{{{\pi }^2}}\over {27}} - {{107\,\ln 2}\over {945}} -
     {{107\,\ln v}\over {945}} \right),
\\
\lefteqn{\eta_{3,3}=
{{1215\,{v^2}}\over {896}} - {{1215\,{v^4}}\over {112}} +
  {{3645\,\pi \,{v^5}}\over {448}} + {{243729\,{v^6}}\over {9856}} -
  {{3645\,\pi \,{v^7}}\over {56}}
}\\
& &
+ {v^8}\,\biggl( {{25037019729}\over {125565440}} -
     {{47385\,\gamma }\over {1568}} + {{3645\,{{\pi }^2}}\over {224}} -
     {{47385\,\ln 2}\over {1568}}
\\
& &
- {{47385\,\ln 3}\over {1568}} -
     {{47385\,\ln v}\over {1568}} \biggr),
\\
\lefteqn{\eta_{3,2}=
{{5\,{v^4}}\over {63}} - {{193\,{v^6}}\over {567}} +
  {{20\,\pi \,{v^7}}\over {63}} + {{86111\,{v^8}}\over {280665}},
}\\
\lefteqn{\eta_{3,1}=
{{{v^2}}\over {8064}} - {{{v^4}}\over {1512}}
+ {{\pi \,{v^5}}\over {4032}} +
  {{437\,{v^6}}\over {266112}} - {{\pi \,{v^7}}\over {756}}
}\\
& &
+ {v^8}\,\biggl( -{{1137077}\over {50854003200}} -
     {{13\,\gamma }\over {42336}} + {{{{\pi }^2}}\over {6048}} -
     {{13\,\ln 2}\over {42336}} - {{13\,\ln v}\over {42336}}
\biggr),
\\
\lefteqn{\eta_{4,4}=
{{1280\,{v^4}}\over {567}} - {{151808\,{v^6}}\over {6237}} +
  {{10240\,\pi \,{v^7}}\over {567}}
+ {{560069632\,{v^8}}\over {6243237}},
}\\
\lefteqn{\eta_{4,3}=
{{729\,{v^6}}\over {4480}} - {{28431\,{v^8}}\over {24640}},
}\\
\lefteqn{\eta_{4,2}=
{{5\,{v^4}}\over {3969}} - {{437\,{v^6}}\over {43659}} +
  {{20\,\pi \,{v^7}}\over {3969}}
+ {{7199152\,{v^8}}\over{218513295}},
}\\
\lefteqn{\eta_{4,1}=
{{{v^6}}\over {282240}} - {{101\,{v^8}}\over {4656960}},
}\\
\lefteqn{\eta_{5,5}=
{{9765625\,{v^6}}\over {2433024}}
 - {{2568359375\,{v^8}}\over {47443968}},
}\\
\lefteqn{\eta_{5,4}=
{{4096\,{v^8}}\over {13365}},
}\\
\lefteqn{\eta_{5,3}=
{{2187\,{v^6}}\over {450560}} - {{150903\,{v^8}}\over {2928640}},
}\\
\lefteqn{\eta_{5,2}=
{{4\,{v^8}}\over {40095}},
}\\
\lefteqn{\eta_{5,1}=
{{{v^6}}\over {127733760}} - {{179\,{v^8}}\over {2490808320}},
}\\
\lefteqn{\eta_{6,6}=
{26244\over 3575}v^8,
}\\
\lefteqn{\eta_{6,4}=
{131072\over 9555975}v^8,
}\\
\lefteqn{\eta_{6,2}=
{4\over 5733585}v^8.
}\\
\end{eqnarray*}
{}From the above results, we obtain $dE/dt$ which is correct up to
$O(v^8)$ as
\begin{eqnarray}
{dE\over dt}&=&
{\left(dE\over dt\right)_N}\Biggl(
1 - {{1247\,{v^2}}\over {336}} + 4\,\pi \,{v^3} -
  {{44711\,{v^4}}\over {9072}} - {{8191\,\pi \,{v^5}}\over {672}} -
  {{16285\,\pi \,{v^7}}\over {504}}
\nonumber\\
& &
+{v^6}\,\biggl( {{6643739519}\over {69854400}} -
     {{1712\,{\gamma}}\over {105}} + {{16\,{{\pi }^2}}\over 3}
- {{3424\,\ln 2}\over {105}} - {{1712\,\ln v}\over {105}} \biggr)
\nonumber\\
& &
+{v^8}\,\biggl( -{{323105549467}\over {3178375200}} +
     {{232597\,{\gamma}}\over {4410}} -
     {{1369\,{{\pi }^2}}\over {126}} + {{39931\,\ln 2}\over {294}}
\nonumber\\
& &
- {{47385\,\ln 3}\over {1568}} + {{232597\,\ln v}\over {4410}}
\biggr)
\Biggr).
\label{eq:dedt}
\end{eqnarray}

\subsection{Wave forms}

Noting the form of the spherical harmonics,
${}_sY_{\ell m}(\theta,\varphi)
=\,{}_sP_{\ell m}(\theta)e^{im\varphi}$, and that of
$A^{\in}_\ell$ given by Eqs.(\ref{eq:ain2}) $\sim$ (\ref{eq:ain}),
we find the wave form is in the form,
\begin{eqnarray}
h^+_{\ell m}-ih^{\times}_{\ell m}&\propto &
e^{i\epsilon(\ln 2\epsilon+\gamma)}e^{-im\Omega(t-r^*)}
e^{im\varphi}
\nonumber\\
&=&e^{im[2v^3(\gamma+2\ln 2+3\ln v)-\Omega(t-r^*)+\varphi]}
e^{2imv^3\ln m}.
\label{eq:defp}
\end{eqnarray}
This suggests that it is more convenient to introduce a
new phase variable $\psi$ as
\begineq
\psi=\Omega(t-r^*)-\varphi-2v^3(\gamma+2\ln 2+3\ln v).
\endeq
An advantage of introducing the phase variable $\psi$ is that we
directly see the post-Newtonian corrections at $O(v^3)$ to
the phase of waves.
These phase corrections at $O(v^3)$ delay the arrival time
of the wave to the observer and express the tail effects.
 From Eqs.(\ref{eq:ain2}) $\sim$ (\ref{eq:ain4}), we also see
the tail effects at $O(v^6)$ on the phase.

In passing, we note that if one expanded $A^{\in}_\ell$ in terms of
$v$, there would appear terms like $(\ln v)^2$ at $O(v^6)$.
Then one would be surprized to see the cancellation of them
 in the calculation of $O(\epsilon^2)$.
However in our approach, we see from the beginning that such
terms do not exist.
Together with the fact that there are no $\ln v$ terms
at orders smaller than $O(v^5)$ and no $(\ln v)^2$ terms at all
in $X^{\in}_\ell$,
we readily foresee that there will be no $(\ln v)^2$ terms
in $dE/dt$ to $O(v^8)$, as we have explicitly seen in the previous
subsection.

Using $\psi$ we obtain
\begin{eqnarray}
\lefteqn{h^{+}_{\ell m}+h^{+}_{\ell,-m}=}
\nonumber\\
& & {\mu\over r}{1\over \omega}
\left(\,_{-2}P_{\ell m}(\theta)
+(-1)^\ell\,_{-2}P_{\ell,-m}(\theta)\right)
\left[W_{\ell m}e^{-im\psi}+{\bar W}_{\ell m}e^{im\psi}\right],
\label{eq:hp}
\end{eqnarray}
\begin{eqnarray}
\lefteqn{h^{\times}_{\ell m}+h^{\times}_{\ell,-m}=}
\nonumber\\
& & {\mu\over r}{i\over \omega}
\left(\,_{-2}P_{\ell m}(\theta)
-(-1)^\ell\,_{-2}P_{\ell,-m}(\theta)\right)
\left[W_{\ell m}e^{-im\psi}-{\bar W}_{\ell m}e^{im\psi}\right],
\label{eq:hx}
\end{eqnarray}
where $W_{\ell m}$ is defined as
\begineq
Z_{\ell m}=W_{\ell m}e^{im\varphi}e^{2v^3(\gamma+2\ln 2+3\ln v)}.
\endeq
Now from Eq.(\ref{eq:zetadef}) and using the explcit forms
of ${}_{-2}P_{\ell m}(\theta)$,
it is straightforward to obtain $\zeta_{\ell m}^{+,\times}$.
The explicit wave form for each $\ell$ and $m$
are given in Appendix B.

\subsection{Comparison with the numerical result}

In this section we compare our analytical result for $dE/dt$
with the numerical result calculated by
Tagoshi and Nakamura \cite{Ref:TN}.

The numerical values of the coefficents in the luminosity formula
calculated in this paper are given in Table 1(a).
\begin{table}[htb]
\hskip5mm
\begin{tabular}{|l|r|} \hline
$v^4$      & $-4.928461199294533$\\ \hline
$v^5$      & $+38.29283545469344$\\ \hline
$v^6$      & $+115.7317166756113$\\ \hline
$v^6\ln v$ & $-16.3047619047619 $\\ \hline
$v^7$      & $-101.5095959597416$\\ \hline
$v^8$      & $-117.5043907226773$\\ \hline
$v^8\ln v$ & $+52.74308390022676$\\ \hline
\end{tabular}
\hskip15mm
\begin{tabular}{|l|r|} \hline
$v^4$      & $-4.928461199295258$ \\ \hline
$v^5$      & $-38.29283545329089$ \\ \hline
$v^6$      & $+115.73172132$      \\ \hline
$v^6\ln v$ & $-16.304761151$      \\ \hline
$v^7$      & $-101.509987$        \\ \hline
$v^8$      & $-117.787$           \\ \hline
$v^8\ln v$ & $+52.6901$           \\ \hline
$v^9$      & $+700.45$            \\ \hline
$v^9\ln v$ & $-209.78$            \\ \hline
\end{tabular}\\
{}~\hfil\\
\hbox{\hskip30mm(a)\hskip70mm (b)\hfil}
\caption{(a) The numerical values of the analytically
calculated coefficients of the post-Newtonian expansion of
$dE/dt$ given in Eq.(\protect\ref{eq:dedt}), and
(b) those numerically calculated by means of least square fitting,
 where the same numerical data as in
Tagoshi and Nakamura \protect\cite{Ref:TN} are used,
but additionally including $\ln v$ terms at $O(v^9)$ and $0(v^{10})$.}
\end{table}
Comparing them with the coefficients given in their paper,
we see that their coefficient of $v^8$ differs from
the true value about $\sim 20\%$, while all the other coefficients
are in very good agreement with each other.
This difference is within the error estimated in their paper.
However, we find that in fact their numerical data turn out to
have much better accuracy. The reason is as follows.
In Ref.\cite{Ref:TN}, the coefficients are calculated
by least square fitting, but taking account of $\ln v$ terms
only at $O(v^6)$ and $O(v^8)$. However, since
they calculated the gravitational waves to $\ell=6$,
their data correctly contain contributions from the orders
 up to $O(v^9)$ of the post-Newtonian expansion and
some constributions from yet higher orders,
 provided the data have enough accuracy.
Hence we have recalculated the coefficients using their
numerical data by least square fitting,
including $v^9\ln v$ and $v^{10}\ln v$ terms.
The result are given in Table 1(b).
We find that even the coefficient of $v^8$ agrees with the
analytical value within $1\%$.
This suggests that their data do actually give the
coefficients to $O(v^9)$ of the post-Newtonian expansion
with high accuracy, hence the values of the coefficients of
$v^9$ and $v^9\ln v$ given in Table 1(b) are expected to be
good approximations to the true values.

\section{An estimate of orbital phase of coalescing binaries}

In this section, based on our results,
 we discuss the accuracy of the post-Newtonian
expansion to construct the template wave forms from inspiraling
compact binaries.
Although our results are valid only for the test particle limit,
$\mu/M\ll1$, we ignore this fact in the following. Since the effect
of non-vanishing $\mu/M$ would only increase errors in the estimate,
our estimate below may be regarded as an optimistic one.

The total cycle $N$ of gravitational waves from an inspiraling compact
binary during sweep through, say, the LIGO band is
\begineq
N=\int^{t_f}_{t_i}fdt
=\int^{r_i}_{r_f}dr{\Omega\over \pi}{{dE/dr}\over {\mid dE/dt\mid}},
\label{eq:cycle}
\endeq
where $f$ is the orbital frequency,
$t_i$($t_f$) and $r_i$($r_f$) are the initial (final) time and
the orbital separation of the binary, respectively,
and we have assumed quasi-periodicity of the inspiral orbit.
Then expanding both the numerator and denominator with respect to
$v$, $N$ is expressed as
\begineq
N={5\over 32\pi}{M\over \mu}\int^{x_f}_{x_i}dx
{\sum_{k=0}^\infty b_kx^k\over{x^6\,\sum_{k=0}^{\infty}a_kx^k}},
\endeq
where $x=(M/r)^{1/2}$ and the series forms in the denominator
and numerator sympolically represent
the post-Newtonian corrections to the $dE/dr$ and $dE/dt$,
respectively, i.e, $\sum b_kx^k=(dE/dr)/(dE/dr)_N$ and
 $\sum a_k x^k=(dE/dt)/(dE/dt)_N$.

To examine the accuracy of the post-Newtonian expansion,
we introduce $N^{(n)}$ which is defined by
\begin{eqnarray}
N^{(n)}={5\over 32\pi}{M\over \mu}\int^{x_f}_{x_i}dx
{{\sum_{k=0}^{n}b_kx^k}\over
{x^6\,\sum_{k=0}^{n}a_kx^k}}.
\label{eq:cyc}
\end{eqnarray}
To be specific, we assume the detectable frequency band to
be from $10\,$Hz to $1000\,$Hz and the quasi-periodic inspiral
stage ends at $r=6M$. So, $r_i$ is the radius at which
$f(r_i)=10\,$Hz, and $r_f$ is the one at which
$f(r_f)=1000\,$Hz if $r_f>6M$ and $r=6M$ otherwise.

For $(1.4M_{\odot},1.4M_{\odot})$ neutron star binary,
we have $r_i=175M$ and $r_f=8M$.
Then we obtain $N^{(5)}=16220.13$, $N^{(6)}=16211.03$,
$N^{(7)}=16211.93$ and $N^{(8)}=16211.91$.
For $(1.4M_{\odot},10M_{\odot})$ binary,
$r_i=68M$ and $r_f=6M$, and we obtain
$N^{(5)}=3484.83$, $N^{(6)}=3466.56$,
$N^{(7)}=3468.60$ and $N^{(8)}=3468.27$.
For $(10M_{\odot},10M_{\odot})$ binary,
$r_i=47M$ and $r_f=6M$, and we obtain
$N^{(5)}=581.39$, $N^{(6)}=574.49$,
$N^{(7)}=575.29$ and $N^{(8)}=575.15$.

 From these results we see that
for $(1.4M_{\odot},1.4M_{\odot})$ binaries,
the post-Newtonian expansion to $O(v^7)$ will be necessary to
accurately predict the total cycle.
However for $(1.4M_{\odot},$ $10M_{\odot})$
and $(10M_{\odot},$ $10M_{\odot})$ binaries,
the convergence seems to be slow.
The error in the total cycle seems marginal, $\Delta N<0.5$,
even including up to $O(v^8)$. Further, if we regard the value of
the coefficient of $v^9$ in Table 1(b) to be approximately correct,
about 700, the contribution from $O(v^9)$
may be comparable to $O(v^8)$.
For these binaries, more detailed analyses are needed.

\section{Conclusion}
Based upon formulae developed in Paper I,
we have analytically perfomed the post-Newtonian expansion
of the gravitational waves from a test particle in circular orbit
around a Schwarzschild black hole.

We have calculated both the gravitational wave forms and luminosity
to order $v^8$ beyond Newtonian. In particular, we have given the
exact analytical values of the coefficients of $\ln v$ terms
at $O(v^6)$ and $O(v^8)$ in the luminosity.
The existence of such terms was found previously in
a numerical analysis by Tagoshi and Nakamura \cite{Ref:TN}.
We have found that their numerical results are in very good
agreement with our analytical ones.

We have also estimated the accuracy of the post-Newtonian
expansion to predict the total cycle of coalescing binaries.
For $(1.4M_{\odot},1.4M_{\odot})$ binaries,
we have found that the post-Newtonian expansion
up to $O(v^7)$ will be sufficient to construct accurate
theoretical templates.
However for $(1.4M_{\odot},10M_{\odot})$ and
$(10M_{\odot},10M_{\odot})$ binaries,
 accuracy to order higher than $v^8$ seems to be necessary.

Although valid only in the test particle
limit, our analytical results should be reproduced
in the conventional post-Newtonian calculations
which are not restricted to the test particle case.
Hence our results give a useful guideline for the future researches
in the gravitational wave physics of coalescing compact binaries.

\begin{center}
{\large ACKNOWLEDGMENTS}
\end{center}

We thank T. Nakamura, M. Shibata and T. Tanaka for useful discussions.
Numerical calculations in this paper are based on the
work of HT with T. Nakamura.
HT thanks Prof. H.Sato for continuous encouragement.
This work is supported by the Grant-in Aid for Scientific
Research on Priority Area of the Ministry of Education
(04234104).

\appendix
\section{The expansion of $X^{\in}_{\ell}$}
In this Appendix, we give the expansion forms of $X^{(n)}_{\ell}$
which are required to calculate $X^{\in}_\ell$ up to order $v^8$.

\begin{eqnarray}
X^{(0)}_2&=&
{{{z^3}}\over {15}} - {{{z^5}}\over {210}} + {{{z^7}}\over {7560}} -
{{{z^9}}\over {498960}} + {{{z^{11}}}\over {51891840}}
\nonumber\\
X^{(1)}_2&=&
{{-13\,{z^4}}\over {630}} + {{{z^6}}\over {810}} -
  {{53\,{z^8}}\over {1782000}} + {{227\,{z^{10}}}\over {567567000}}
\nonumber\\
X^{(2)}_2&=&
{z^3}\,\left( {{26743}\over {110250}} - {{107\,\ln z}\over {3150}} \right)
+ {z^5}\,\left( -{{140953}\over {9261000}} +
     {{107\,\ln z}\over {44100}} \right)
\nonumber\\
X^{(3)}_2&=&
{{319\,{z^2}}\over {6300}}
\nonumber\\
X^{(0)}_3&=&
{{{z^4}}\over {105}} - {{{z^6}}\over {1890}} + {{{z^8}}\over {83160}} -
  {{{z^{10}}}\over {6486480}}
\nonumber\\
X^{(1)}_3&=&
{{-{z^3}}\over {126}} - {{{z^5}}\over {630}} +
  {{221\,{z^7}}\over {2494800}}
\nonumber\\
X^{(2)}_3&=&
{{76369\,{z^3}}\over {1852200}} - {{2327663\,{z^5}}\over {1100206800}}
- {{13\,{z^3}\,\ln z}\over {4410}} +
  {{13\,{z^5}\,\ln z}\over {79380}}
\nonumber\\
X^{(0)}_4&=&
{{{z^5}}\over {945}} - {{{z^7}}\over {20790}}
+ {{{z^9}}\over {1081080}} - {{{z^{11}}}\over {97297200}}
\nonumber\\
X^{(1)}_4&=&
{{-{z^4}}\over {630}} - {{{z^6}}\over {9900}} +
  {{79\,{z^8}}\over {13899600}}
\nonumber\\
X^{(2)}_4&=&
{{{z^3}}\over {1764}}
 + {{{z^5}\, \left( 942578261
 - 43548120\,\ln z \right) }\over {181534122000}}
\nonumber\\
X^{(0)}_5&=&
{{{z^6}}\over {10395}} - {{{z^8}}\over {270270}} +
  {{{z^{10}}}\over {16216200}}
\nonumber\\
X^{(1)}_5&=&
{{-{z^5}}\over {4950}} - {{41\,{z^7}}\over {8108100}}
\nonumber\\
X^{(2)}_5&=&{{{z^4}}\over {7425}}
\nonumber\\
X^{(0)}_6&=&
{{{z^7}}\over {135135}} - {{{z^9}}\over {4054050}} +
  {{{z^{11}}}\over {275675400}}
\nonumber\\
X^{(1)}_6&=&
{{-8\,{z^6}}\over {405405}} - {{{z^8}}\over {5675670}}
\nonumber\\
X^{(2)}_6&=&
{{4\,{z^5}}\over {212355}}
\nonumber\\
X^{(0)}_7&=&
{{{z^8}}\over {2027025}} - {{{z^{10}}}\over {68918850}}
\nonumber\\
X^{(1)}_7&=&
{{-{z^7}}\over {630630}}
\nonumber\\
X^{(0)}_8&=&
{{{z^9}}\over {34459425}} - {{{z^{11}}}\over {1309458150}}
\nonumber\\
X^{(1)}_8&=&
{{-{z^8}}\over {9189180}}
\nonumber\\
X^{(0)}_9&=&
{{{z^{10}}}\over {654729075}}
\nonumber\\
X^{(0)}_{10}&=&{{{z^{11}}}\over {13749310575}}
\nonumber
\end{eqnarray}

\section{Wave forms to $O(v^8)$}
In this Appendix, we give the gravitational wave forms
for all the relevant $\ell$ and $m$ that contribute up to $O(v^8)$.

\begin{eqnarray*}
\\
\lefteqn{\zeta^{+}_{2,2}=
-\left( 3 + \cos (2\,\theta ) \right) \,
  \left( \cos (2\,\psi ) - {{107\,{v^2}\,\cos (2\,\psi )}\over {42}}
\right.}
\\
& &
+ {v^3}\,\left( 2\,\pi \,\cos (2\,\psi ) +
         \left( -{{17}\over 3} + 4\,\ln 2 \right) \,\sin (2\,\psi ) \right)
  -   {{2173\,{v^4}\,\cos (2\,\psi )}\over {1512}}
\\
& &
  +   {v^5}\,\left( {{-107\,\pi \,\cos (2\,\psi )}\over {21}} +
   \left( {{1819}\over {126}} - {{214\,\ln 2}\over {21}} \right) \,
   \sin (2\,\psi ) \right)
\\
& &
+ {v^6}\,\left( \cos (2\,\psi )\,
          \left( {{49928027}\over {1940400}} - {{856\,\gamma }\over {105}} +
            {{2\,{{\pi }^2}}\over 3} + {{668\,\ln 2}\over {105}} -
     8\,{{(\ln 2)}^2} - {{856\,\ln v}\over {105}} \right)  \right.
\\
& &
\left.
\quad+ \left( {{-254\,\pi }\over {35}} + 8\,\pi \,\ln 2 \right) \,
          \sin (2\,\psi ) \right)
\\
& &
   + {v^7}\,\left( {{-2173\,\pi \,\cos (2\,\psi )}\over {756}} +
    \left( {{36941}\over {4536}} - {{2173\,\ln 2}\over {378}} \right)
    \,\sin (2\,\psi ) \right)
\\
& &
 + {v^8}\,\left( \cos (2\,\psi )\,
    \left( -{{326531600453}\over {12713500800}} +
       {{45796\,\gamma }\over {2205}} - {{107\,{{\pi }^2}}\over {63}}
\right.
\right.
\\
& &
\left.
\quad
- {{35738\,\ln 2}\over {2205}} +
       {{428\,{{(\ln 2)}^2}}\over {21}} +
       {{45796\,\ln v}\over {2205}} \right)
\\
& &
\left.
\left.
\quad + \left( {{13589\,\pi }\over {735}} -
    {{428\,\pi \,\ln 2}\over {21}} \right) \,\sin (2\,\psi ) \right)
\right),
\\
\lefteqn{\zeta^{+}_{2,1}=
{{4\,\sin (\theta )}\over 3}
\,\left( v\,\sin (\psi ) - {{17\,{v^3}\,\sin (\psi )}\over {28}}
+ {v^4}\,\left( {{10\,\cos (\psi )}\over 3} + \pi \,\sin (\psi ) \right)
- {{43\,{v^5}\,\sin (\psi )}\over {126}}
\right.}
\nonumber\\
& &
+ {v^6}\,\left( {{-85\,\cos (\psi )}\over {42}} -
          {{17\,\pi \,\sin (\psi )}\over {28}} \right)
\nonumber\\
& &
+ {v^7}\,\left( {{81\,\pi \,\cos (\psi )}\over {35}}
+ \left( {{14641367}\over {2910600}} - {{214\,\gamma }\over {105}} +
             {{{{\pi }^2}}\over 6} - {{214\,\ln 2}\over {105}} -
             {{214\,\ln v}\over {105}} \right) \,\sin (\psi ) \right)
\nonumber\\
& &
\left.
+ {v^8}\,\left( {{-215\,\cos (\psi )}\over {189}} -
          {{43\,\pi \,\sin (\psi )}\over {126}} \right)
             \right),
\nonumber\\
\lefteqn{\zeta^{+}_{3,3}=
{{9\,\left( 5\,\sin (\theta ) + \sin (3\,\theta ) \right) }\over {16}}
\,\left( v\,\sin (3\,\psi ) - 4\,{v^3}\,\sin (3\,\psi )
\right.}
\nonumber\\
& &
+ {v^4}\,\left( \cos (3\,\psi )\,
           \left( {{127}\over {10}} - 6\,\ln 3 \right)  +
          3\,\pi \,\sin (3\,\psi ) \right)
+ {{123\,{v^5}\,\sin (3\,\psi )}\over {110}}
\nonumber\\
& &
+ {v^6}\,\left( \cos (3\,\psi )\,
           \left( -{{254}\over 5} + 24\,\ln 3 \right)  -
          12\,\pi \,\sin (3\,\psi ) \right)
\nonumber\\
& &
+ {v^7}\,\left( \cos (3\,\psi )\,
           \left( {{2277\,\pi }\over {70}} - 18\,\pi \,\ln 3 \right)  +
          \left( -{{185741}\over {70070}} - {{78\,\gamma }\over 7} +
\right.
\right.
\nonumber\\
& &
\left.
\left.
\quad     {{3\,{{\pi }^2}}\over 2} - {{78\,\ln 2}\over 7} +
             {{2277\,\ln 3}\over {35}} - 18\,{{(\ln 3)}^2} -
             {{78\,\ln v}\over 7} \right) \,\sin (3\,\psi ) \right)
\nonumber\\
& &
\left.
+ {v^8}\,\left( \cos (3\,\psi )\,
      \left( {{15621}\over {1100}} - {{369\,\ln 3}\over {55}} \right)
     + {{369\,\pi \,\sin (3\,\psi )}\over {110}} \right)
\right)
\nonumber\\
\lefteqn{\zeta^{+}_{3,2}=
{{-4}\over 3}\,\cos (2\,\theta )\,\left( {v^2}\,\cos (2\,\psi ) -
       {{193\,{v^4}\,\cos (2\,\psi )}\over {90}}
\right.}
\nonumber\\
& &
+ {v^5}\,\left( 2\,\pi \,\cos (2\,\psi ) +
          \left( -{{26}\over 3} + 4\,\ln 2 \right) \,\sin (2\,\psi )
          \right)
- {{1451\,{v^6}\,\cos (2\,\psi )}\over {3960}}
\nonumber\\
& &
+ {v^7}\,\left( {{-193\,\pi \,\cos (2\,\psi )}\over {45}} +
          \left( {{2509}\over {135}} - {{386\,\ln 2}\over {45}} \right) \,
           \sin (2\,\psi ) \right)
\nonumber\\
& &
+ {v^8}\,\left( \cos (2\,\psi )\,
           \left( -{{340998173}\over {75675600}} -
             {{104\,\gamma }\over {21}} + {{2\,{{\pi }^2}}\over 3} +
             {{520\,\ln 2}\over {21}} - 8\,{{(\ln 2)}^2} -
\right.
\right.
\nonumber\\
& &
\left.
\left.
\left.
\quad    {{104\,\ln v}\over {21}} \right)  +
          \left( {{-104\,\pi }\over 7} + 8\,\pi \,\ln 2 \right) \,
           \sin (2\,\psi ) \right)  \right)
\nonumber\\
\lefteqn{\zeta^{+}_{3,1}=
{{\left( -\sin (\theta ) + 3\,\sin (3\,\theta ) \right) }\over {48}}
\left( v\,\sin (\psi ) - {{8\,{v^3}\,\sin (\psi )}\over 3}
\right.}
\nonumber\\
& &
+ {v^4}\,\left( {{127\,\cos (\psi )}\over {30}} + \pi \,\sin (\psi )
   \right)
+  {{607\,{v^5}\,\sin (\psi )}\over {198}}
\nonumber\\
& &
+  {v^6}\,\left( {{-508\,\cos (\psi )}\over {45}} -
          {{8\,\pi \,\sin (\psi )}\over 3} \right)
\nonumber\\
& &
+ {v^7}\,\left( {{253\,\pi \,\cos (\psi )}\over {70}} +
      \left( -{{1656587}\over {1891890}} - {{26\,\gamma }\over {21}} +
        {{{{\pi }^2}}\over 6} - {{26\,\ln 2}\over {21}} -
        {{26\,\ln v}\over {21}} \right) \,\sin (\psi ) \right)
\nonumber\\
& &
\left.
+ {v^8}\,\left( {{77089\,\cos (\psi )}\over {5940}} +
          {{607\,\pi \,\sin (\psi )}\over {198}} \right)
\right)
\nonumber\\
\lefteqn{\zeta^{+}_{4,4}=
{{1}\over 3}\left( 5 - 4\,\cos (2\,\theta ) - \cos (4\,\theta ) \right) \,
\left( {v^2}\,\cos (4\,\psi ) -
       {{593\,{v^4}\,\cos (4\,\psi )}\over {110}}
\right.}
\nonumber\\
& &
+ {v^5}\,\left( 4\,\pi \,\cos (4\,\psi ) +
          \left( -{{296}\over {15}} + 8\,\ln 4 \right) \,\sin (4\,\psi )
          \right)
+ {{1068671\,{v^6}\,\cos (4\,\psi )}\over {200200}}
\nonumber\\
& &
+ {v^7}\,\left( {{-1186\,\pi \,\cos (4\,\psi )}\over {55}} +
          \left( {{87764}\over {825}} - {{2372\,\ln 4}\over {55}} \right)\,
           \sin (4\,\psi ) \right)
\nonumber\\
& &
+ {v^8}\,\left( \cos (4\,\psi )\,
           \left( -{{36840955871}\over {499458960}} -
             {{50272\,\gamma }\over {3465}} + {{8\,{{\pi }^2}}\over 3} -
             {{50272\,\ln 2}\over {3465}}
             + {{496736\,\ln 4}\over {3465}}
\right.
\right.
\nonumber\\
& &
\left.
\left.
\left.
     - 32\,{{(\ln 4)}^2} -
             {{50272\,\ln v}\over {3465}} \right)  +
          \left( {{-248368\,\pi }\over {3465}} + 32\,\pi \,\ln 4 \right) \,
           \sin (4\,\psi ) \right)  \right)
\nonumber\\
\lefteqn{\zeta^{+}_{4,3}=
{{-27}\over {80}}\,\left( \sin (\theta ) - 3\,\sin (3\,\theta ) \right)
\,\left( {v^3}\,\sin (3\,\psi ) -
       {{39\,{v^5}\,\sin (3\,\psi )}\over {11}}
\right.}
\nonumber\\
& &
+ {v^6}\,\left( \cos (3\,\psi )\,
           \left( {{149}\over {10}} - 6\,\ln 3 \right)  +
          3\,\pi \,\sin (3\,\psi ) \right)
+ {{7206\,{v^7}\,\sin (3\,\psi )}\over {5005}}
\nonumber\\
& &
\left.
+ {v^8}\,\left( \cos (3\,\psi )\,
           \left( -{{5811}\over {110}} + {{234\,\ln 3}\over {11}} \right)
           - {{117\,\pi \,\sin (3\,\psi )}\over {11}} \right)
\right)
\nonumber\\
\lefteqn{\zeta^{+}_{4,2}=
{{-1}\over {84}}
\left( 5 + 4\,\cos (2\,\theta ) + 7\,\cos (4\,\theta ) \right) \,
\left( {v^2}\,\cos (2\,\psi ) -
         {{437\,{v^4}\,\cos (2\,\psi )}\over {110}}
\right.}
\nonumber\\
& &
+ {v^5}\,\left( 2\,\pi \,\cos (2\,\psi ) +
            \left( -{{148}\over {15}} + 4\,\ln 2 \right) \,\sin (2\,\psi )
            \right)
+ {{1038039\,{v^6}\,\cos (2\,\psi )}\over {200200}}
\nonumber\\
& &
+ {v^7}\,\left( {{-437\,\pi \,\cos (2\,\psi )}\over {55}} +
            \left( {{32338}\over {825}} - {{874\,\ln 2}\over {55}}
\right)\,\sin (2\,\psi ) \right)
\nonumber\\
& &
+ {v^8}\,\left( \cos (2\,\psi )\,
             \left( -{{54548715967}\over {2497294800}} -
               {{12568\,\gamma }\over {3465}} + {{2\,{{\pi }^2}}\over 3} +
               {{111616\,\ln 2}\over {3465}} - 8\,{{(\ln 2)}^2} -
\right.
\right.
\nonumber\\
& &
\left.
\left.
\left.
     {{12568\,\ln v}\over {3465}} \right)  +
   \left( {{-62092\,\pi }\over {3465}} + 8\,\pi \,\ln 2 \right) \,
      \sin (2\,\psi ) \right)
\right)
\nonumber\\
\lefteqn{\zeta^{+}_{4,1}=
{{\left( 3\,\sin (\theta ) + 7\,\sin (3\,\theta ) \right) }\over {560}}
\left( {v^3}\,\sin (\psi ) - {{101\,{v^5}\,\sin (\psi )}\over {33}}
+ {v^6}\,\left( {{149\,\cos (\psi )}\over {30}} + \pi \,\sin (\psi )
   \right)
\right.}
\nonumber\\
& &
+ {{42982\,{v^7}\,\sin (\psi )}\over {15015}}
\left.
+ {v^8}\,\left( {{-15049\,\cos (\psi )}\over {990}} -
          {{101\,\pi \,\sin (\psi )}\over {33}} \right)
\right)
\nonumber\\
\end{eqnarray*}
\begin{eqnarray*}
\\
\lefteqn{\zeta^{+}_{5,5}=
{{625\left( -14\,\sin (\theta ) + 3\,\sin (3\,\theta ) +
       \sin (5\,\theta ) \right) }\over {3072}}
\,\left( {v^3}\,\sin (5\,\psi ) -
       {{263\,{v^5}\,\sin (5\,\psi )}\over {39}}
\right.}
\nonumber\\
& &
+ {v^6}\,\left( \cos (5\,\psi )\,
           \left( {{569}\over {21}} - 10\,\ln 5 \right)  +
          5\,\pi \,\sin (5\,\psi ) \right)
+ {{9185\,{v^7}\,\sin (5\,\psi )}\over {819}}
\nonumber\\
& &
\left.
+ {v^8}\,\left( \cos (5\,\psi )\,
           \left( -{{149647}\over {819}} + {{2630\,\ln 5}\over {39}} \right
             )  - {{1315\,\pi \,\sin (5\,\psi )}\over {39}} \right)
\right)
\nonumber\\
\lefteqn{\zeta^{+}_{5,4}=
{{32\,\left( \cos (2\,\theta ) - \cos (4\,\theta ) \right)}\over {45}}
\,\left( {v^4}\,\cos (4\,\psi )
-{{4451\,{v^6}\,\cos (4\,\psi )}\over {910}}
\right.}
\nonumber\\
& &
\left.
+ {v^7}\,\left( 4\,\pi \,\cos (4\,\psi ) +
   \left( -{{326}\over {15}} + 8\,\ln 4 \right) \,\sin (4\,\psi )
   \right)
+ {{10715\,{v^8}\,\cos (4\,\psi )}\over {2184}}
\right)
\nonumber\\
\lefteqn{\zeta^{+}_{5,3}=
{{27\left( 14\,\sin (\theta ) + 13\,\sin (3\,\theta ) +
       15\,\sin (5\,\theta ) \right) }\over {5120}}
\,\left( {v^3}\,\sin (3\,\psi )
- {{69\,{v^5}\,\sin (3\,\psi )}\over {13}}
\right.}
\nonumber\\
& &
+{v^6}\,\left( \cos (3\,\psi )\,
           \left( {{569}\over {35}} - 6\,\ln 3 \right)  +
          3\,\pi \,\sin (3\,\psi ) \right)
+ {{12463\,{v^7}\,\sin (3\,\psi )}\over {1365}}
\nonumber\\
& &
\left.
+ {v^8}\,\left( \cos (3\,\psi )\,
      \left( -{{39261}\over {455}} + {{414\,\ln 3}\over {13}} \right)
      - {{207\,\pi \,\sin (3\,\psi )}\over {13}} \right)
\right)
\nonumber\\
\lefteqn{\zeta^{+}_{5,2}=
{{-2\,\left( \cos (2\,\theta ) + 3\,\cos (4\,\theta ) \right)}\over {135}} \,
\left( {v^4}\,\cos (2\,\psi ) -
       {{3911\,{v^6}\,\cos (2\,\psi )}\over {910}}
\right.}
\nonumber\\
& &
\left.
+ {v^7}\,\left( 2\,\pi \,\cos (2\,\psi ) +
  \left( -{{163}\over {15}} + 4\,\ln 2 \right) \,\sin (2\,\psi )
  \right)
+ {{63439\,{v^8}\,\cos (2\,\psi )}\over {10920}}
\right)
\nonumber\\
\lefteqn{\zeta^{+}_{5,1}=
{{\left( -2\,\sin (\theta ) - 3\,\sin (3\,\theta ) +
       15\,\sin (5\,\theta ) \right) }\over {23040}}
\left( {v^3}\,\sin (\psi ) - {{179\,{v^5}\,\sin (\psi )}\over {39}}
\right.}
\nonumber\\
& &
+ {v^6}\,\left( {{569\,\cos (\psi )}\over {105}} +
          \pi \,\sin (\psi ) \right)
+ {{5023\,{v^7}\,\sin (\psi )}\over {585}}
\nonumber\\
& &
\left.
+ {v^8}\,\left( {{-101851\,\cos (\psi )}\over {4095}} -
          {{179\,\pi \,\sin (\psi )}\over {39}} \right)
\right)
\nonumber
\end{eqnarray*}
\begin{eqnarray*}
\nonumber\\
\lefteqn{\zeta^{+}_{6,6}=
{{-81\,\left( 3 + \cos (2\,\theta ) \right)\sin^4 (\theta )}\over {40}}
\left( {v^4}\,\cos (6\,\psi ) -
       {{113\,{v^6}\,\cos (6\,\psi )}\over {14}}
\right. }
\nonumber\\
& &
\left.
+ {v^7}\,\left( 6\,\pi \,\cos (6\,\psi ) +
  \left( -{{487}\over {14}} + 12\,\ln 6 \right) \,\sin (6\,\psi )
  \right)
+ {{1372317\,{v^8}\,\cos (6\,\psi )}\over {73304}}
\right),
\nonumber\\
\lefteqn{\zeta^{+}_{6,5}=
{{3125\,\left( 1 - 5\,{{\cos (\theta )}^2} \right)
\sin^3(\theta )}\over {2016}}
\left( {v^5}\,\sin (5\,\psi ) -
       {{149\,{v^7}\,\sin (5\,\psi )}\over {24}}
\right.}
\nonumber\\
& &
\left.
+ {v^8}\,\left( \cos (5\,\psi )\,
           \left( {{1219}\over {42}} - 10\,\ln 5 \right)  +
          5\,\pi \,\sin (5\,\psi ) \right)  \right)
\nonumber
\\
\lefteqn{\zeta^{+}_{6,4}=
{{16\,{{\cos^2 ({{\theta }\over 2})}}\,
{{\sin^2 ({{\theta }\over 2})}}}\over {495}}
\left( 67 + 92\,\cos (2\,\theta ) + 33\,\cos (4\,\theta ) \right) }
\nonumber\\
& &
\left( {v^4}\,\cos (4\,\psi ) -
       {{93\,{v^6}\,\cos (4\,\psi )}\over {14}}
\right.
\nonumber\\
& &
\left.
+ {v^7}\,\left( 4\,\pi \,\cos (4\,\psi ) +
  \left( -{{487}\over {21}} + 8\,\ln 4 \right) \,\sin (4\,\psi ) \right)
+ {{3261767\,{v^8}\,\cos (4\,\psi )}\over {219912}}
    \right)
\nonumber
\\
\lefteqn{\zeta^{+}_{6,3}=
{{243\,\left( 21 + 52\,\cos (2\,\theta ) + 55\,\cos (4\,\theta ) \right)
\sin (\theta )}\over {98560}} }
\nonumber\\
& &
\,\left( {v^5}\,\sin (3\,\psi ) -
       {{133\,{v^7}\,\sin (3\,\psi )}\over {24}}
\right.
\nonumber\\
& &
\left.
+ {v^8}\,\left( \cos (3\,\psi )\,
           \left( {{1219}\over {70}} - 6\,\ln 3 \right)  +
          3\,\pi \,\sin (3\,\psi ) \right)  \right)
\nonumber
\\
\lefteqn{\zeta^{+}_{6,2}=
{{-1}\over {190080}}
\left( 210 + 289\,\cos (2\,\theta ) + 30\,\cos (4\,\theta ) +
         495\,\cos (6\,\theta ) \right) }
\nonumber\\
& &
\left( {v^4}\,\cos (2\,\psi ) -
         {{81\,{v^6}\,\cos (2\,\psi )}\over {14}}
+ {v^7}\,\left( 2\,\pi \,\cos (2\,\psi )
+ \left( -{{487}\over {42}} + 4\,\ln 2 \right) \,\sin (2\,\psi )
            \right)
\right.
\nonumber\\
& &
\left.
+ {{14482483\,{v^8}\,\cos (2\,\psi )}\over {1099560}}
\right)
\nonumber\\
\lefteqn{\zeta^{+}_{6,1}=
{{\left( 35 + 60\,\cos (2\,\theta ) + 33\,\cos (4\,\theta ) \right) \,
\sin (\theta )}\over {266112}} }
\nonumber\\
& &
\left( {v^5}\,\sin (\psi ) - {{125\,{v^7}\,\sin (\psi )}\over {24}}
+ {v^8}\,\left( {{1219\,\cos (\psi )}\over {210}} +
          \pi \,\sin (\psi ) \right)
\right)
\nonumber\\
\end{eqnarray*}
\begin{eqnarray*}
\lefteqn{\zeta^{+}_{7,7}=
{{117649\,\left( 3 + \cos (2\,\theta ) \right) \,
\sin^5 (\theta )}\over {46080}}
\left( {v^5}\,\sin (7\,\psi )
-{{319\,{v^7}\,\sin (7\,\psi )}\over {34}}
\right.}
\nonumber\\
& &
\left.
+{v^8}\,\left( \cos (7\,\psi )\,
\left( {{7699}\over {180}} - 14\,\ln 7 \right)  +
7\,\pi \,\sin (7\,\psi ) \right)  \right)
\nonumber\\
\lefteqn{\zeta^{+}_{7,6}=
{{-243\left( 2 + 3\,\cos (2\,\theta ) \right) \,\sin^4 (\theta )}
\over {140}}
}\nonumber\\
& &
\left( {v^6}\,\cos (6\,\psi ) -
{{1787\,{v^8}\,\cos (6\,\psi )}\over {238}} \right)
\nonumber\\
\lefteqn{\zeta^{+}_{7,5}=
{{-15625\,\left( 233 + 316\,\cos (2\,\theta )
+ 91\,\cos (4\,\theta ) \right){{\sin^3 (\theta )}}}\over {3354624}}
\biggl( {v^5}\,\sin (5\,\psi )}
\nonumber\\
& &
- {{271\,{v^7}\,\sin (5\,\psi )}\over {34}}
+ {v^8}\,\left( \cos (5\,\psi )\,
\left( {{7699}\over {252}} - 10\,\ln 5 \right)  +
5\,\pi \,\sin (5\,\psi ) \right)  \biggr)
\nonumber\\
\lefteqn{\zeta^{+}_{7,4}=
{{32\left( 9 + 18\,\cos (2\,\theta ) + 13\,\cos (4\,\theta ) \right)
{{\sin^2 (\theta )}}}\over {1365}}
}\nonumber\\
& &
\,\left( {v^6}\,\cos (4\,\psi )
- {{14543\,{v^8}\,\cos (4\,\psi )}\over {2142}} \right)
\nonumber\\
\lefteqn{\zeta^{+}_{7,3}=
{{729\sin (\theta )}\over {82001920}}
\,\bigl( 1134 + 1863\,\cos (2\,\theta ) + 1122\,\cos (4\,\theta )
+ 1001\,\cos (6\,\theta ) \bigr)}
\nonumber\\
& &
\left( {v^5}\,\sin (3\,\psi )
 -{{239\,{v^7}\,\sin (3\,\psi )}\over {34}}
 + {v^8}\,\left( \cos (3\,\psi )\,
\left( {{7699}\over {420}} - 6\,\ln 3 \right)  +
          3\,\pi \,\sin (3\,\psi ) \right)  \right)
\nonumber\\
\lefteqn{\zeta^{+}_{7,2}=
{{-1}\over {192192}}
\left( 25\,\cos (2\,\theta ) + 88\,\cos (4\,\theta ) +
         143\,\cos (6\,\theta ) \right)
}\nonumber\\
& &
\left( {v^6}\,\cos (2\,\psi )
- {{13619\,{v^8}\,\cos (2\,\psi )}\over {2142}} \right)
\nonumber\\
\lefteqn{\zeta^{+}_{7,1}=
{{\left( 350 + 775\,\cos (2\,\theta ) + 946\,\cos (4\,\theta ) +
       1001\,\cos (6\,\theta ) \right)\sin (\theta )}\over {147603456}}
}\nonumber\\
& &
\left( {v^5}\,\sin (\psi )
- {{223\,{v^7}\,\sin (\psi )}\over {34}}
+ {v^8}\,\left( {{7699\,\cos (\psi )}\over {1260}} +
          \pi \,\sin (\psi ) \right)  \right)
\nonumber\\
\lefteqn{\zeta^{+}_{8,8}=
{{1024\,{{\sin^6 (\theta )}}}\over {315}}
\,\left( {v^6}\,\cos (8\,\psi ) -
       {{3653\,{v^8}\,\cos (8\,\psi )}\over {342}} \right) \,
     \left( 3 + \cos (2\,\theta ) \right)
}\nonumber\\
\lefteqn{\zeta^{+}_{8,7}=
{{-823543\,{v^7}\,\left( 5 + 7\,\cos (2\,\theta ) \right) \,\sin (7\,\psi )\,
     {{\sin^5 (\theta )}}}\over {829440}}
}\nonumber\\
\lefteqn{\zeta^{+}_{8,6}=
{{-729 \cos^4 (\theta )\,\sin^4 (\theta )}\over {140}}
\,\left( {v^6}\,\cos (6\,\psi ) -
       {{353\,{v^8}\,\cos (6\,\psi )}\over {38}} \right) \,
}\nonumber\\
\lefteqn{\zeta^{+}_{8,5}=
{-78125\over {4644864}}
\left( 33 + 60\,\cos (2\,\theta ) +
35\,\cos (4\,\theta ) \right) \,{{\sin^3 (\theta )}}
\,v^7\sin (5\,\psi )
}\nonumber\\
\lefteqn{\zeta^{+}_{8,4}=
{{\left( 186 + 309\,\cos (2\,\theta ) + 182\,\cos (4\,\theta ) +
       91\,\cos (6\,\theta ) \right) \,{{\sin^2 (\theta )}}}\over {4095}}
}\nonumber\\
& &
\left( {v^6}\,\cos (4\,\psi ) -
{{2837\,{v^8}\,\cos (4\,\psi )}\over {342}} \right) \,
\nonumber\\
\lefteqn{\zeta^{+}_{8,3}=
{{-243}\over {7454720}}
{v^7}\sin (3\,\psi )
\,\left( 1 + 3\,\cos (2\,\theta ) \right) \,
\left( 15 - 52\,\cos (2\,\theta ) - 91\,\cos (4\,\theta ) \right) \,
     \,\sin (\theta )
}\nonumber\\
\lefteqn{\zeta^{+}_{8,2}=
{{-\left( 315 + 512\,\cos (2\,\theta ) + 220\,\cos (4\,\theta ) +
1001\,\cos (8\,\theta ) \right)
}\over {23063040}}
}\nonumber\\
& &
\left( {v^6}\,\cos (2\,\psi ) -
{{2633\,{v^8}\,\cos (2\,\psi )}\over {342}} \right) \,
\nonumber\\
\lefteqn{\zeta^{+}_{8,1}=
{{v^7}\over {379551744}}
\,\left( 210 + 385\,\cos (2\,\theta ) + 286\,\cos (4\,\theta ) +
       143\,\cos (6\,\theta ) \right) \,\sin (\psi )\,\sin (\theta )
}\nonumber\\
\lefteqn{\zeta^{+}_{9,9}=
{{-4782969\,{v^7}}\over {1146880}}
\,\left( 3 + \cos (2\,\theta ) \right) \,\sin (9\,\psi )\,
     {{\sin^7 (\theta )}}
}\nonumber\\
\lefteqn{\zeta^{+}_{9,8}=
{{{32768\,{v^8}}}\over {14175}}
\,\cos (8\,\psi )\,\left( 3 + 4\,\cos (2\,\theta ) \right) \,
     {{\sin^6 (\theta )}}
}\nonumber\\
\lefteqn{\zeta^{+}_{9,7}=
{{5764801\,{v^7}}\over {902430720}}\,\sin (7\,\psi )
\,\left( 515 + 676\,\cos (2\,\theta ) +
       153\,\cos (4\,\theta ) \right) \,{{\sin^5 (\theta )}}
}\nonumber\\
\lefteqn{\zeta^{+}_{9,6}=
{{-729\,{v^8}\,\cos (6\,\psi )\,}\over {23800}}
     \left( 39 + 67\,\cos (2\,\theta ) + 34\,\cos (4\,\theta ) \right) \,
     {{\sin^4 (\theta )}}
}\nonumber\\
\lefteqn{\zeta^{+}_{9,5}=
{{-390625\,{v^7}}\over {1263403008}}\sin (5\,\psi )
\left( 462 + 759\,\cos (2\,\theta ) +
\right.}
\nonumber\\
& &
\left.
418\,\cos (4\,\theta ) + 153\,\cos (6\,\theta ) \right)\,
{{\sin^3 (\theta )}}
\nonumber\\
\lefteqn{\zeta^{+}_{9,4}=
{{16\,{v^8}}\over {34425}}
\,\cos (4\,\psi )\,\left( 33 + 66\,\cos (2\,\theta ) +
\right.}
\nonumber\\
& &
\left.
59\,\cos (4\,\theta ) + 34\,\cos (6\,\theta ) \right) \,
     {{\sin^2 (\theta )}}
\nonumber\\
\lefteqn{\zeta^{+}_{9,3}=
{{243\,{v^7}}\over {579338240}}\sin (3\,\psi )
\left( 693 + 1232\,\cos (2\,\theta ) + 884\,\cos (4\,\theta )
\right.}
\nonumber\\
& &
\left.
+ 624\,\cos (6\,\theta ) + 663\,\cos (8\,\theta ) \right) \,
\sin (\theta )
\nonumber\\
\lefteqn{\zeta^{+}_{9,2}=
{{-v^8}\over {57283200}}\,\cos (2\,\psi )\,
\left( 49\,\cos (2\,\theta ) + 182\,\cos (4\,\theta )
\right.}
\nonumber\\
& &
\left.
+ 351\,\cos (6\,\theta ) + 442\,\cos (8\,\theta ) \right)
\nonumber\\
\lefteqn{\zeta^{+}_{9,1}=
{{{v^7}}\over {93852794880}}\sin (\psi )
\,\left( 735 + 1568\,\cos (2\,\theta ) + 1820\,\cos (4\,\theta )
\right.}
\nonumber\\
& &
\left.
+ 2080\,\cos (6\,\theta ) + 1989\,\cos (8\,\theta ) \right) \,
\sin (\theta )
\nonumber\\
\lefteqn{\zeta^{+}_{10,10}=
{{-390625\,{v^8}}\over {72576}}
\,\cos (10\,\psi )\,\left( 3 + \cos (2\,\theta ) \right) \,
     {{\sin^8 (\theta )}}
}\nonumber\\
\lefteqn{\zeta^{+}_{10,8}=
{{4096\,{v^8}}\over {269325}}
\,\cos (8\,\psi )\,
\left( 349 + 452\,\cos (2\,\theta ) + 95\,\cos (4\,\theta ) \right) \,
{{\sin^6 (\theta )}}
}\nonumber\\
\lefteqn{\zeta^{+}_{10,6}=
{{-4374\,{v^8}}\over {231017675}}
\,\cos (6\,\psi )\,
\left( 19318 + 31299\,\cos (2\,\theta )
\right.}
\nonumber\\
& &
\left.
+ 16218\,\cos (4\,\theta ) +
4845\,\cos (6\,\theta ) \right) \,{{\sin^4 (\theta )}}
\nonumber\\
\lefteqn{\zeta^{+}_{10,4}=
{{{v^8}}\over {4578525}}\,\cos (4\,\psi )
\,\left( 10791 + 19272\,\cos (2\,\theta ) +
13868\,\cos (4\,\theta )
\right.}
\nonumber\\
& &
\left.
+ 8568\,\cos (6\,\theta ) +
4845\,\cos (8\,\theta ) \right) \,{{\sin^2 (\theta )}}
\nonumber\\
\lefteqn{\zeta^{+}_{10,2}=
{{-{v^8}}\over {139312742400}}
\,\cos (2\,\psi )\,
\left( 16170 + 28322\,\cos (2\,\theta ) + 17576\,\cos (4\,\theta )
\right.}
\nonumber\\
& &
\left.
+ 4693\,\cos (6\,\theta ) + 1326\,\cos (8\,\theta ) +
62985\,\cos (10\,\theta ) \right)
\nonumber
\end{eqnarray*}

\begin{eqnarray*}
\\
\lefteqn{\zeta^{\times}_{2,2}=
4\,\cos (\theta )\,
\left( \sin (2\,\psi ) -
    {{107\,{v^2}\,\sin (2\,\psi )}\over {42}}
+ {v^3}\,\left( \cos (2\,\psi )\,
   \left( {{17}\over 3} - 4\,\log (2) \right)  + 2\,\pi \,\sin (2\,\psi )
   \right)
\right.}
\\
& &
- {{2173\,{v^4}\,\sin (2\,\psi )}\over {1512}}
+ {v^5}\,\left( \cos (2\,\psi )\,
    \left( -{{1819}\over {126}} + {{214\,\log (2)}\over {21}} \right)  -
   {{107\,\pi \,\sin (2\,\psi )}\over {21}} \right)
\\
& &
+ {v^6}\,\left( \cos (2\,\psi )\,
      \left( {{254\,\pi }\over {35}} - 8\,\pi \,\log (2) \right)
\right.
\\
& &
\left.
+  \left( {{49928027}\over {1940400}} - {{856\,\gamma }\over {105}} +
          {{2\,{{\pi }^2}}\over 3} + {{668\,\log (2)}\over {105}} -
          8\,{{\log (2)}^2} - {{856\,\log (v)}\over {105}} \right) \,
        \sin (2\,\psi ) \right)
\\
& &
+ {v^7}\,\left( \cos (2\,\psi )\,
    \left( -{{36941}\over {4536}} + {{2173\,\log (2)}\over {378}} \right)
    - {{2173\,\pi \,\sin (2\,\psi )}\over {756}} \right)
\\
& &
+ {v^8}\,\left( \cos (2\,\psi )\,
   \left( {{-13589\,\pi }\over {735}} +
          {{428\,\pi \,\log (2)}\over {21}} \right)  +
   \left( -{{326531600453}\over {12713500800}} +
          {{45796\,\gamma }\over {2205}}
\right.
\right.
\\
& &
\left.
\left.
\left.
- {{107\,{{\pi }^2}}\over {63}} -
          {{35738\,\log (2)}\over {2205}} +
  {{428\,{{\log (2)}^2}}\over {21}} + {{45796\,\log (v)}\over {2205}}
          \right) \,\sin (2\,\psi ) \right)
\right)
\\
\lefteqn{\zeta^{\times}_{2,1}=
{{2\sin (2\,\theta )}\over 3}
\,\left( v\,\cos (\psi ) - {{17\,{v^3}\,\cos (\psi )}\over {28}}
+ {v^4}\,\left( \pi \,\cos (\psi ) - {{10\,\sin (\psi )}\over 3} \right)
\right.}
\\
& &
- {{43\,{v^5}\,\cos (\psi )}\over {126}}
+ {v^6}\,\left( {{-17\,\pi \,\cos (\psi )}\over {28}} +
          {{85\,\sin (\psi )}\over {42}} \right)
+ {v^7}\,\left( \cos (\psi )\,
      \left( {{14641367}\over {2910600}} - {{214\,\gamma }\over {105}}
\right.
\right.
\\
& &
\left.
\left.
+ {{{{\pi }^2}}\over 6} - {{214\,\ln 2}\over {105}} -
      {{214\,\ln v}\over {105}} \right)  -
   {{81\,\pi \,\sin (\psi )}\over {35}} \right)
\\
& &
\left.
+ {v^8}\,\left( {{-43\,\pi \,\cos (\psi )}\over {126}} +
          {{215\,\sin (\psi )}\over {189}} \right)
\right)
\\
\lefteqn{\zeta^{\times}_{3,3}=
{{9\,\sin (2\,\theta )}\over 4}
\,\left( v\,\cos (3\,\psi ) - 4\,{v^3}\,\cos (3\,\psi )
\right.}
\\
& &
+ {v^4}\,\left( 3\,\pi \,\cos (3\,\psi ) +
  \left( -{{127}\over {10}} + 6\,\ln 3 \right) \,\sin (3\,\psi )
  \right)
+ {{123\,{v^5}\,\cos (3\,\psi )}\over {110}}
\\
& &
+ {v^6}\,\left( -12\,\pi \,\cos (3\,\psi )
+ \left( {{254}\over 5} - 24\,\ln 3 \right) \,\sin (3\,\psi )
  \right)
\\
& &
+ {v^7}\,\left( \cos (3\,\psi )\,
           \left( -{{185741}\over {70070}} - {{78\,\gamma }\over 7} +
             {{3\,{{\pi }^2}}\over 2} - {{78\,\ln 2}\over 7} +
             {{2277\,\ln 3}\over {35}} - 18\,{{(\ln 3)}^2}
\right.
\right.
\\
& &
\left.
\left.
- {{78\,\ln v}\over 7} \right)  +
          \left( {{-2277\,\pi }\over {70}} + 18\,\pi \,\ln 3
          \right) \,\sin (3\,\psi ) \right)
\\
& &
\left.
+ {v^8}\,\left( {{369\,\pi \,\cos (3\,\psi )}\over {110}} +
    \left( -{{15621}\over {1100}} + {{369\,\ln 3}\over {55}}
    \right)\,\sin (3\,\psi ) \right)
\right)
\\
\lefteqn{\zeta^{\times}_{3,2}=
{{\left( 5\,\cos (\theta ) + 3\,\cos (3\,\theta ) \right)}\over 6}
\left( {v^2}\,\sin (2\,\psi ) -
       {{193\,{v^4}\,\sin (2\,\psi )}\over {90}}
\right.}
\\
& &
+ {v^5}\,\left( \cos (2\,\psi )\,
           \left( {{26}\over 3} - 4\,\ln 2 \right)  +
          2\,\pi \,\sin (2\,\psi ) \right)
- {{1451\,{v^6}\,\sin (2\,\psi )}\over {3960}}
\\
& &
+ {v^7}\,\left( \cos (2\,\psi )\,
           \left( -{{2509}\over {135}} + {{386\,\ln 2}\over {45}} \right)
           - {{193\,\pi \,\sin (2\,\psi )}\over {45}} \right)
\\
& &
+ {v^8}\,\left( \cos (2\,\psi )\,
     \left( {{104\,\pi }\over 7} - 8\,\pi \,\ln 2 \right)  +
\right.
\\
& &
\left.
\left.
     \left( -{{340998173}\over {75675600}} -
     {{104\,\gamma }\over {21}} + {{2\,{{\pi }^2}}\over 3} +
     {{520\,\ln 2}\over {21}} - 8\,{{(\ln 2)}^2} -
     {{104\,\ln v}\over {21}} \right) \,\sin (2\,\psi ) \right)
\right)
\\
\lefteqn{\zeta^{\times}_{3,1}=
{{\sin (2\,\theta )}\over {12}}
\left( v\,\cos (\psi ) - {{8\,{v^3}\,\cos (\psi )}\over 3}
\right.}
\\
& &
+ {v^4}\,\left( \pi \,\cos (\psi )
      - {{127\,\sin (\psi )}\over {30}} \right)
+ {{607\,{v^5}\,\cos (\psi )}\over {198}}
+ {v^6}\,\left( {{-8\,\pi \,\cos (\psi )}\over 3} +
          {{508\,\sin (\psi )}\over {45}} \right)
\\
& &
+ {v^7}\,\left( \cos (\psi )\,
     \left( -{{1656587}\over {1891890}} - {{26\,\gamma }\over {21}} +
       {{{{\pi }^2}}\over 6} - {{26\,\ln 2}\over {21}} -
        {{26\,\ln v}\over {21}} \right)  -
     {{253\,\pi \,\sin (\psi )}\over {70}} \right)
\\
& &
\left.
+ {v^8}\,\left( {{607\,\pi \,\cos (\psi )}\over {198}} -
          {{77089\,\sin (\psi )}\over {5940}} \right)
\right)
\\
\lefteqn{\zeta^{\times}_{4,4}=
{{-4\,\left( \cos (\theta ) - \cos (3\,\theta ) \right)}\over 3}
\left( {v^2}\,\sin (4\,\psi ) -
   {{593\,{v^4}\,\sin (4\,\psi )}\over {110}}
\right.}
\\
& &
+ {v^5}\,\left( \cos (4\,\psi )\,
           \left( {{296}\over {15}} - 8\,\ln 4 \right)  +
          4\,\pi \,\sin (4\,\psi ) \right)
+ {{1068671\,{v^6}\,\sin (4\,\psi )}\over {200200}}
\\
& &
+ {v^7}\,\left( \cos (4\,\psi )\,
    \left( -{{87764}\over {825}} + {{2372\,\ln 4}\over {55}} \right)
       - {{1186\,\pi \,\sin (4\,\psi )}\over {55}} \right)
\\
& &
+ {v^8}\,\left( \cos (4\,\psi )\,
     \left( {{248368\,\pi }\over {3465}} - 32\,\pi \,\ln 4 \right)  +
     \left( -{{36840955871}\over {499458960}} -
       {{50272\,\gamma }\over {3465}} + {{8\,{{\pi }^2}}\over 3}
\right.
\right.
\\
& &
\left.
\left.
\left.
- {{50272\,\ln 2}\over {3465}} +
       {{496736\,\ln 4}\over {3465}} - 32\,{{(\ln 4)}^2} -
       {{50272\,\ln v}\over {3465}} \right) \,\sin (4\,\psi ) \right)
\right)
\\
\lefteqn{\zeta^{\times}_{4,3}=
{{27\left( 2\,\sin (2\,\theta ) +
       \sin (4\,\theta ) \right) }\over {80}}
\,\left( {v^3}\,\cos (3\,\psi ) -
       {{39\,{v^5}\,\cos (3\,\psi )}\over {11}}
\right.}
\\
& &
+ {v^6}\,\left( 3\,\pi \,\cos (3\,\psi ) +
   \left( -{{149}\over {10}} + 6\,\ln 3 \right) \,\sin (3\,\psi )
   \right)
\\
& &
\left.
+ {{7206\,{v^7}\,\cos (3\,\psi )}\over {5005}}
+ {v^8}\,\left( {{-117\,\pi \,\cos (3\,\psi )}\over {11}} +
     \left( {{5811}\over {110}} - {{234\,\ln 3}\over {11}} \right) \,
      \sin (3\,\psi ) \right)
\right)
\\
\lefteqn{\zeta^{\times}_{4,2}=
{{\left( \cos (\theta ) + 7\,\cos (3\,\theta ) \right)}\over {42}}
\left( {v^2}\,\sin (2\,\psi ) -
       {{437\,{v^4}\,\sin (2\,\psi )}\over {110}}
\right.}
\\
& &
+ {v^5}\,\left( \cos (2\,\psi )\,
       \left( {{148}\over {15}} - 4\,\ln 2 \right)  +
          2\,\pi \,\sin (2\,\psi ) \right)
+ {{1038039\,{v^6}\,\sin (2\,\psi )}\over {200200}}
\\
& &
+ {v^7}\,\left( \cos (2\,\psi )\,
  \left( -{{32338}\over {825}} + {{874\,\ln 2}\over {55}} \right)
   - {{437\,\pi \,\sin (2\,\psi )}\over {55}} \right)
\\
& &
+ {v^8}\,\left( \cos (2\,\psi )\,
    \left( {{62092\,\pi }\over {3465}} - 8\,\pi \,\ln 2 \right)  +
    \left( -{{54548715967}\over {2497294800}} -
\right.
\right.
\\
& &
\left.
\left.
\left.
      {{12568\,\gamma }\over {3465}} + {{2\,{{\pi }^2}}\over 3} +
      {{111616\,\ln 2}\over {3465}} - 8\,{{(\ln 2)}^2} -
      {{12568\,\ln v}\over {3465}} \right) \,\sin (2\,\psi ) \right)
\right)
\\
\lefteqn{\zeta^{\times}_{4,1}=
{{\left( -2\,\sin (2\,\theta ) + 7\,\sin (4\,\theta ) \right) }\over {560}}
\left( {v^3}\,\cos (\psi ) - {{101\,{v^5}\,\cos (\psi )}\over {33}}
\right.}
\\
& &
+ {{42982\,{v^7}\,\cos (\psi )}\over {15015}}
+ {v^6}\,\left( \pi \,\cos (\psi ) - {{149\,\sin (\psi )}\over {30}}
   \right)
\\
& &
\left.
+ {v^8}\,\left( {{-101\,\pi \,\cos (\psi )}\over {33}} +
          {{15049\,\sin (\psi )}\over {990}} \right)
\right)
\\
\end{eqnarray*}
\begin{eqnarray*}
\\
\lefteqn{\zeta^{\times}_{5,5}=
{{625\left( -2\,\sin (2\,\theta ) +
       \sin (4\,\theta ) \right) }\over {768}}
\,\left( {v^3}\,\cos (5\,\psi ) -
       {{263\,{v^5}\,\cos (5\,\psi )}\over {39}}
\right.}
\\
& &
+ {v^6}\,\left( 5\,\pi \,\cos (5\,\psi ) +
   \left( -{{569}\over {21}} + 10\,\ln 5 \right) \,\sin (5\,\psi )
   \right)
+ {{9185\,{v^7}\,\cos (5\,\psi )}\over {819}}
\\
& &
\left.
+ {v^8}\,\left( {{-1315\,\pi \,\cos (5\,\psi )}\over {39}} +
  \left( {{149647}\over {819}} - {{2630\,\ln 5}\over {39}} \right)
    \sin (5\,\psi ) \right)
\right)
\\
\lefteqn{\zeta^{\times}_{5,4}=
{{-2\,\left( 14\,\cos (\theta ) - 9\,\cos (3\,\theta ) -
       5\,\cos (5\,\theta ) \right)}\over {45}}
\left( {v^4}\,\sin (4\,\psi )
- {{4451\,{v^6}\,\sin (4\,\psi )}\over {910}}
\right.}
\\
& &
\left.
+ {v^7}\,\left( \cos (4\,\psi )\,
   \left( {{326}\over {15}} - 8\,\ln 4 \right)  +
  4\,\pi \,\sin (4\,\psi ) \right)
+ {{10715\,{v^8}\,\sin (4\,\psi )}\over {2184}}
\right)
\\
\lefteqn{\zeta^{\times}_{5,3}=
{{-27\,\left( 2\,\sin (2\,\theta )
  - 9\,\sin (4\,\theta ) \right) }\over {1280}}
\,\left( {v^3}\,\cos (3\,\psi ) -
  {{69\,{v^5}\,\cos (3\,\psi )}\over {13}}
\right.}
\\
& &
+ {v^6}\,\left( 3\,\pi \,\cos (3\,\psi ) +
  \left( -{{569}\over {35}} + 6\,\ln 3 \right) \,\sin (3\,\psi )
  \right)
+ {{12463\,{v^7}\,\cos (3\,\psi )}\over {1365}}
\\
& &
\left.
+ {v^8}\,\left( {{-207\,\pi \,\cos (3\,\psi )}\over {13}} +
  \left( {{39261}\over {455}} - {{414\,\ln 3}\over {13}} \right) \,
  \sin (3\,\psi ) \right)
\right)
\\
\lefteqn{\zeta^{\times}_{5,2}=
{{\left( 14\,\cos (\theta ) + 3\,\cos (3\,\theta ) +
       15\,\cos (5\,\theta ) \right)}\over {540}}
\left( {v^4}\,\sin (2\,\psi )
- {{3911\,{v^6}\,\sin (2\,\psi )}\over {910}}
\right.}
\\
& &
\left.
+ {v^7}\,\left( \cos (2\,\psi )\,
           \left( {{163}\over {15}} - 4\,\ln 2 \right)  +
          2\,\pi \,\sin (2\,\psi ) \right)
+ {{63439\,{v^8}\,\sin (2\,\psi )}\over {10920}}
\right)
\\
\lefteqn{\zeta^{\times}_{5,1}=
{{\left( 2\,\sin (2\,\theta ) + 3\,\sin (4\,\theta )
  \right) }\over {5760}}
\left( {v^3}\,\cos (\psi ) - {{179\,{v^5}\,\cos (\psi )}\over {39}}
\right.}
\\
& &
+ {v^6}\,\left( \pi \,\cos (\psi ) -
          {{569\,\sin (\psi )}\over {105}} \right)
+ {{5023\,{v^7}\,\cos (\psi )}\over {585}}
\\
& &
\left.
+ {v^8}\,\left( {{-179\,\pi \,\cos (\psi )}\over {39}} +
    {{101851\,\sin (\psi )}\over {4095}} \right)
\right)
\\
\lefteqn{\zeta^{\times}_{6,6}=
{{81\,\cos (\theta )\,\sin^4 (\theta )}\over {10}}
\,\left( {v^4}\,\sin (6\,\psi ) -
       {{113\,{v^6}\,\sin (6\,\psi )}\over {14}}
\right.}
\\
& &
\left.
+ {v^7}\,\left( \cos (6\,\psi )\,
       \left( {{487}\over {14}} - 12\,\ln 6 \right)  +
          6\,\pi \,\sin (6\,\psi ) \right)
+ {{1372317\,{v^8}\,\sin (6\,\psi )}\over {73304}}
\right)
\\
\lefteqn{\zeta^{\times}_{6,5}=
{{-3125\,\cos (\theta )\,\left( 1 + 3\,{{\cos^2 (\theta )}} \right) \,
{{\sin^3 (\theta )}}}\over {2016}}
\left( {v^5}\,\cos (5\,\psi ) -
       {{149\,{v^7}\,\cos (5\,\psi )}\over {24}}
\right.}
\\
& &
\left.
+ {v^8}\,\left( 5\,\pi \,\cos (5\,\psi ) +
   \left( -{{1219}\over {42}} + 10\,\ln 5 \right) \,\sin (5\,\psi )
          \right)
\right)
\\
\lefteqn{\zeta^{\times}_{6,4}=
{{-64\,\cos (\theta )\,\left( 1 + 11\,\cos (2\,\theta ) \right) \,
\sin^2 (\theta )}\over {495}}
\left( {v^4}\,\sin (4\,\psi ) -
       {{93\,{v^6}\,\sin (4\,\psi )}\over {14}}
\right.}
\\
& &
\left.
+ {v^7}\,\left( \cos (4\,\psi )\,
           \left( {{487}\over {21}} - 8\,\ln 4 \right)  +
          4\,\pi \,\sin (4\,\psi ) \right)
+ {{3261767\,{v^8}\,\sin (4\,\psi )}\over {219912}}
\right)
\\
\lefteqn{\zeta^{\times}_{6,3}=
{{243\,\cos (\theta )\,\sin (\theta )\,
\left( 53 + 20\,\cos (2\,\theta ) +
       55\,\cos (4\,\theta ) \right) }\over {98560}}
\left( {v^5}\,\cos (3\,\psi )
\right.}
\\
& &
\left.
-{{133\,{v^7}\,\cos (3\,\psi )}\over {24}}
+ {v^8}\,\left( 3\,\pi \,\cos (3\,\psi ) +
  \left( -{{1219}\over {70}} + 6\,\ln 3 \right) \,\sin (3\,\psi )
  \right)
\right)
\\
\lefteqn{\zeta^{\times}_{6,2}=
{{\cos (\theta )\,\left( 47 - 84\,\cos (2\,\theta ) +
       165\,\cos (4\,\theta ) \right)}\over {23760}}
\left( {v^4}\,\sin (2\,\psi )
\right.}
\\
& &
- {{81\,{v^6}\,\sin (2\,\psi )}\over {14}}
+ {v^7}\,\left( \cos (2\,\psi )\,
      \left( {{487}\over {42}} - 4\,\ln 2 \right)
\right.
\\
& &
\left.
\left.
+ 2\,\pi \,\sin (2\,\psi ) \right)
+ {{14482483\,{v^8}\,\sin (2\,\psi )}\over {1099560}}
\right)
\\
\lefteqn{\zeta^{\times}_{6,1}=
{{\cos (\theta )\,\sin (\theta )\,\left( 41 - 12\,\cos (2\,\theta ) +
       99\,\cos (4\,\theta ) \right)}\over {266112}}
}\\
& &
\left( {v^5}\,\cos (\psi ) - {{125\,{v^7}\,\cos (\psi )}\over {24}} +
       {v^8}\,\left( \pi \,\cos (\psi ) -
          {{1219\,\sin (\psi )}\over {210}} \right)
\right)
\\
\end{eqnarray*}
\begin{eqnarray*}
\\
\lefteqn{\zeta^{\times}_{7,7}=
{{117649\,\cos (\theta )\,\sin^5 (\theta )}\over {11520}}
\,\left( {v^5}\,\cos (7\,\psi ) -
       {{319\,{v^7}\,\cos (7\,\psi )}\over {34}}
\right.}
\\
& &
\left.
+ {v^8}\,\left( 7\,\pi \,\cos (7\,\psi ) +
  \left( -{{7699}\over {180}} + 14\,\ln 7 \right) \,\sin (7\,\psi )
           \right)
\right)
\\
\lefteqn{\zeta^{\times}_{7,6}=
{{243\,\cos (\theta ){{\sin^4 (\theta )}}
\,\left( 13 + 7\,\cos (2\,\theta ) \right)}\over {560}}
}\\
& &
\left( {v^6}\,\sin (6\,\psi ) -
       {{1787\,{v^8}\,\sin (6\,\psi )}\over {238}}
\right)
\\
\lefteqn{\zeta^{\times}_{7,5}=
{{-78125\,\cos (\theta )\sin^3 (\theta )
\,\left( 3 + 13\,\cos (2\,\theta ) \right)}\over {419328}}
\left( {v^5}\,\cos (5\,\psi )
\right.}
\\
& &
\left.
- {{271\,{v^7}\,\cos (5\,\psi )}\over {34}}
+ {v^8}\,\left( 5\,\pi \,\cos (5\,\psi ) +
   \left( -{{7699}\over {252}} + 10\,\ln 5 \right) \,\sin (5\,\psi )
   \right)
\right)
\\
\lefteqn{\zeta^{\times}_{7,4}=
{{-4\,\cos (\theta )\,\left( 113 + 116\,\cos (2\,\theta ) +
       91\,\cos (4\,\theta ) \right) \sin^2 (\theta )}\over {1365}}
}\\
& &
\left( {v^6}\,\sin (4\,\psi ) -
       {{14543\,{v^8}\,\sin (4\,\psi )}\over {2142}} \right)
\\
\lefteqn{\zeta^{\times}_{7,3}=
{{-729\,\cos (\theta )\,\left( 167 + 44\,\cos (2\,\theta ) +
  429\,\cos (4\,\theta ) \right) \,\sin (\theta )}\over {10250240}}
}\\
& &
\left( {v^5}\,\cos (3\,\psi ) -
       {{239\,{v^7}\,\cos (3\,\psi )}\over {34}}
\right.
\\
& &
\left.
+ {v^8}\,\left( 3\,\pi \,\cos (3\,\psi ) +
    \left( -{{7699}\over {420}} + 6\,\ln 3 \right) \,\sin (3\,\psi )
        \right)
\right)
\\
\lefteqn{\zeta^{\times}_{7,2}=
{{\cos (\theta )\,\left( -338 + 1351\,\cos (2\,\theta ) -
 990\,\cos (4\,\theta ) + 1001\,\cos (6\,\theta ) \right) }\over {768768}}
}\\
& &
\left( {v^6}\,\sin (2\,\psi ) -
  {{13619\,{v^8}\,\sin (2\,\psi )}\over {2142}} \right)
\\
\lefteqn{\zeta^{\times}_{7,1}=
{{\cos (\theta )\,\sin (\theta )\,\left( 109 + 132\,\cos (2\,\theta ) +
       143\,\cos (4\,\theta ) \right)}\over {18450432}}
}\\
& &
\left( {v^5}\,\cos (\psi ) - {{223\,{v^7}\,\cos (\psi )}\over {34}} +
       {v^8}\,\left( \pi \,\cos (\psi ) -
          {{7699\,\sin (\psi )}\over {1260}} \right)
\right)
\\
\lefteqn{\zeta^{\times}_{8,8}=
{{-4096\,\cos (\theta )\,{{\sin^6 (\theta )}}}\over {315}}
\left( {v^6}\,\sin (8\,\psi ) -
       {{3653\,{v^8}\,\sin (8\,\psi )}\over {342}}
\right)
}\\
\lefteqn{\zeta^{\times}_{8,7}=
{{-823543\,{v^7}}\over {207360}}\,\cos (7\,\psi )
\,\left( 2 + \cos (2\,\theta ) \right) \,{{\sin^5 (\theta )}}
}\\
\lefteqn{\zeta^{\times}_{8,6}=
{{729\,\cos (\theta )\,\left( 1 + 3\,\cos (2\,\theta ) \right) \,
{{\sin^4 (\theta )}}}\over {560}}
}\\
& &
\left( {v^6}\,\sin (6\,\psi ) -
       {{353\,{v^8}\,\sin (6\,\psi )}\over {38}}
\right)
\\
\lefteqn{\zeta^{\times}_{8,5}=
{{-78125\,{v^7}}\over {1161216}}
\,\cos (5\,\psi )\,\cos (\theta )\,
\left( 11 + 14\,\cos (2\,\theta ) + 7\,\cos (4\,\theta ) \right) \,
     {{\sin^3 (\theta )}}
}\\
\lefteqn{\zeta^{\times}_{8,4}=
{{-4\,\cos (\theta )\,\left( 49 + 52\,\cos (2\,\theta ) +
91\,\cos (4\,\theta ) \right) \,{{\sin^2 (\theta )}}}\over {4095}}
}\\
& &
\left( {v^6}\,\sin (4\,\psi ) -
       {{2837\,{v^8}\,\sin (4\,\psi )}\over {342}} \right) \,
\\
\lefteqn{\zeta^{\times}_{8,3}=
{{243\,{v^7}}\over {3727360}}\,\cos (3\,\psi )\,\cos (\theta )\,
     \left( 24 + 141\,\cos (2\,\theta ) + 91\,\cos (6\,\theta ) \right) \,
     \sin (\theta )
}\\
\lefteqn{\zeta^{\times}_{8,2}=
{{\cos (\theta )\,\left( -274 + 583\,\cos (2\,\theta ) -
  286\,\cos (4\,\theta ) + 1001\,\cos (6\,\theta ) \right)}\over {11531520}}
}\\
& &
\left( {v^6}\,\sin (2\,\psi ) -
  {{2633\,{v^8}\,\sin (2\,\psi )}\over {342}} \right)
\\
\lefteqn{\zeta^{\times}_{8,1}=
{{{v^7}}\over {94887936}}\,\cos (\psi )\,\cos (\theta )\,
\left( -8 + 121\,\cos (2\,\theta ) + 143\,\cos (6\,\theta ) \right) \,
\sin (\theta )
}\\
\lefteqn{\zeta^{\times}_{9,9}=
{{-4782969\,{v^7}}\over{286720}}\,\cos (9\,\psi )
\,\cos (\theta )\,{{\sin^7 (\theta )}}
}\\
\lefteqn{\zeta^{\times}_{9,8}=
{{-8192\,{v^8}}\over {14175}}\,\sin (8\,\psi )
\,\cos (\theta )\,\left( 19 + 9\,\cos (2\,\theta ) \right) \,
{{\sin^6 (\theta )}}
}\\
\lefteqn{\zeta^{\times}_{9,7}=
{{40353607\,{v^7}}\over {112803840}}
\,\cos (7\,\psi )\,\cos (\theta )\,
\left( 7 + 17\,\cos (2\,\theta ) \right) \,{{\sin^5 (\theta )}}
}\\
\lefteqn{\zeta^{\times}_{9,6}=
{{729\,{v^8}\,\sin (6\,\psi )}\over {47600}}
\,\cos (\theta )\,\left( 97 + 132\,\cos (2\,\theta ) +
51\,\cos (4\,\theta ) \right) \,{{\sin^4 (\theta )}}
}\\
\lefteqn{\zeta^{\times}_{9,5}=
{{-390625\,{v^7}}\over {157925376}}
\,\cos (5\,\psi )\,\cos (\theta )\,
\left( 59 + 80\,\cos (2\,\theta ) + 85\,\cos (4\,\theta ) \right) \,
{{\sin^3 (\theta )}}
}\\
\lefteqn{\zeta^{\times}_{9,4}=
{{-4\,{v^8}}\over {34425}}\sin (4\,\psi )
\,\cos (\theta )\,\left( 142 + 343\,\cos (2\,\theta )
\right.}
\\
& &
\left.
+ 130\,\cos (4\,\theta ) + 153\,\cos (6\,\theta ) \right) \,
     \,{{\sin^2 (\theta )}}
\\
\lefteqn{\zeta^{\times}_{9,3}=
{{243\,{v^7}}\over {72417280}}
\,\cos (3\,\psi )\,\cos (\theta )\,
  \left( 18 + 195\,\cos (2\,\theta )
\right.}
\\
& &
\left.
+ 78\,\cos (4\,\theta ) +
   221\,\cos (6\,\theta ) \right) \,\sin (\theta )
\\
\lefteqn{\zeta^{\times}_{9,2}=
{{{v^8}}\over {114566400}}\,\sin (2\,\psi )
\,\cos (\theta )\,\left( 1279 - 1480\,\cos (2\,\theta ) +
\right.}\\
& &
\left.
   2236\,\cos (4\,\theta ) - 1976\,\cos (6\,\theta ) +
   1989\,\cos (8\,\theta ) \right)
\\
\lefteqn{\zeta^{\times}_{9,1}=
{{{v^7}}\over {11731599360}}
\,\cos (\psi )\,\cos (\theta )\,
  \left( 166 + 403\,\cos (2\,\theta ) + 234\,\cos (4\,\theta )
\right.}
\\
& &
\left.
+ 221\,\cos (6\,\theta ) \right) \,\sin (\theta )
\\
\lefteqn{\zeta^{\times}_{10,10}=
{{390625\,{v^8}\,\sin (10\,\psi )}\over {18144}}
\,\cos (\theta )\,{{\sin^8 (\theta )}}
}\\
\lefteqn{\zeta^{\times}_{10,8}=
{{-131072\,{v^8}\sin (8\,\psi )}\over {269325}}
\,\cos (\theta )\,\left( 9 + 19\,\cos (2\,\theta ) \right) \,
     \,{{\sin^6 (\theta )}}
}\\
\lefteqn{\zeta^{\times}_{10,6}=
{{2187\,{v^8}\,\sin (6\,\psi )}\over {14470400}}
\,\cos (\theta )\,\left( 2449 + 3604\,\cos (2\,\theta ) +
 2907\,\cos (4\,\theta ) \right) \,{{\sin^4 (\theta )}}
}\\
\lefteqn{\zeta^{\times}_{10,4}=
{{-16\,{v^8}\sin (4\,\psi )}\over {4578525}}
\,\cos (\theta )\,\left( 446 + 1319\,\cos (2\,\theta )
\right.}
\\
& &
\left.
+ 850\,\cos (4\,\theta ) + 969\,\cos (6\,\theta ) \right) \,
\,{{\sin^2 (\theta )}}
\\
\lefteqn{\zeta^{\times}_{10,2}=
{{{v^8}\,\sin (2\,\psi )}\over {17414092800}}
\,\cos (\theta )\,\left( 3007 - 5720\,\cos (2\,\theta )
\right.}
\\
& &
\left.
+ 8268\,\cos (4\,\theta ) - 1768\,\cos (6\,\theta ) +
  12597\,\cos (8\,\theta ) \right)
\\
\end{eqnarray*}

\end{document}